\documentclass[10pt]{article}
\usepackage{euscript,amssymb,amsmath}
\usepackage[dvips]{graphicx}

\usepackage{amsmath}
\usepackage{amssymb}
\usepackage{graphicx,xcolor}
\usepackage{subfigure}
\usepackage{adjustbox}
\usepackage{multirow}
\usepackage{authblk}

\usepackage{epstopdf}
\catcode`\@=11

\newcommand{\be}[1]{\begin{equation}\label{#1}}
\newcommand{\ee}{\end{equation}}
\newcommand{\ba}[1]{\begin{eqnarray}\label{#1}}
\newcommand{\ea}{\end{eqnarray}}
\newcommand{\rf}[1]{(\ref{#1})}
\newcommand{\nn}{\nonumber}

\begin{document}

\title{Localizing EP sets in dissipative systems and the \\ self-stability of bicycles}
\date{}
\author{Oleg N. Kirillov\thanks{\textit{Corresponding author.} Northumbria University, Newcastle upon Tyne, NE1 8ST, United Kingdom \\ Email: oleg.kirillov@northumbria.ac.uk }}
\maketitle

\abstract{Sets in the parameter space corresponding to complex exceptional points have high codimension and by this reason they are difficult objects for numerical localization. However, complex EPs play an important role in the problems of stability of dissipative systems where they are frequently considered as precursors to instability. We propose to localize the set of complex EPs using the fact that the minimum of the spectral abscissa of a polynomial is attained at the EP of the highest possible order. Applying this approach to the problem of self-stabilization of a bicycle we find explicitly the EP sets that suggest scaling laws for the design of robust bikes that agree with the design of the known experimental machines.}

\section{Introduction}

Exceptional points in classical systems recently attracted attention of researchers in the context of the parity-time (PT) symmetry found
in mechanics \cite{B2013, Xu2015} and electronics \cite{S2012}. In the context of stability of classical systems the PT-symmetry
plays a part in the systems of mechanical oscillators with the indefinite matrix of damping forces \cite{F1996, FZ1996, FGK1997, F1999, KP2009}.
Stable PT-symmetric indefinitely damped mechanical systems have imaginary eigenvalues and thus form singularities on the boundary of the domain of asymptotic stability of general dissipative systems \cite{K2012,K2013t}. Among these singularities are exceptional points corresponding to double imaginary eigenvalues with the Jordan block. They belong to a set of complex exceptional points with nonzero real parts that lives both in the domain of instability and in the domain of asymptotic stability of a dissipative system and passes through the imaginary exceptional point on the stability boundary that bounds the inerval of PT-symmetry \cite{K2013f,K2017}. This is a set of high codimension which is hard to find by numerical approaches. Nevertheless, in many applications it was realized that complex exceptional points hidden within the domain of asymptotic stability significantly influence the transition to instability \cite{Jones1988,Dobson2001}. How to localize the set of complex exceptional points?
The general approach involving commutators of matrices of the system \cite{F1983, C1999} does not look easily interpretable. In this paper we will use a recent observation \cite{KO2013} that the set of complex exceptional points connects the imaginary exceptional points on the boundary of asymptotic stability and the real exceptional points inside the domain of asymptotic stability that lie on the boundary of the domain of heavy damping. We will show how localization of the exceptional points with this approach helps to find explicit scaling laws in the classical problem of self-stability of bicycles.

\section{Complex exceptional points and the self-stability of bicycles}

Bicycle is easy to ride but surprisingly difficult to model. Refinement of the mathematical model of a bicycle
continues over the last 150 years with contributions from Rankine, Boussinesq, Whipple, Klein, Sommerfeld, Appel, Synge and many others \cite{Ruina2007,Sharp2008}.
A canonical, commonly accepted nowadays model goes back to the 1899 work by Whipple. The Whipple bike is a system consisting of four rigid bodies with knife-edge wheels making it non-holonomic, i.e. requiring for its description more configuration coordinates than the number of its admissible
velocities \cite{Whipple2018,BM2017}. Due to the non-holonomic constraints even the bicycle tire tracks have a nontrivial and beautiful geometry that has deep and unexpected links to integrable systems, particle traps, and the Berry phase \cite{LT2009,L2017,L2018}.

A fundamental empirical property of real bicycles is their self-stability without any control at a sufficiently high speed \cite{TMS2011}.
Understanding the passive stabilization is expected to play a crucial part in formulating principles of design of energy-efficient wheeled and bipedal robots \cite{Bipedal2005}.
However, the theoretical explanation of self-stability has been highly debated throughout the history of bicycle dynamics \cite{Whipple2018}.
The reason to why ``simple questions about self-stabilization of bicycles do not have straightforward answers'' \cite{Sharp2008} lies in the symbolical complexity of the Whipple model that contains 7 degrees of freedom and depends on 25 physical and design parameters \cite{Ruina2007}.
In recent numerical simulations \cite{Ruina2007,Sharp2008,Whipple2018} self-stabilization has been observed for some benchmark designs of the Whipple bike. These results suggested further simplification of the model yielding a reduced model of a bicycle with vanishing radii of the wheels (that are replaced by skates, see e.g. \cite{BM2015}), known as the two-mass-skate (TMS) bicycle \cite{TMS2011}. Despite the self-stable TMS bike has been successfully realized in the recent laboratory experiments \cite{TMS2011}, the reasons for its self-stability still wait for a theoretical explanation.

In the following, we will show how localization of complex and real exceptional points allows to find hidden symmetries in the model suggesting further reduction of the parameter space and, finally, providing explicit relations between the parameters of stability-optimized TMS bikes.

\subsection{The TMS bicycle model}
\begin{figure}
    \begin{center}
    \includegraphics[angle=0, width=0.55\textwidth]{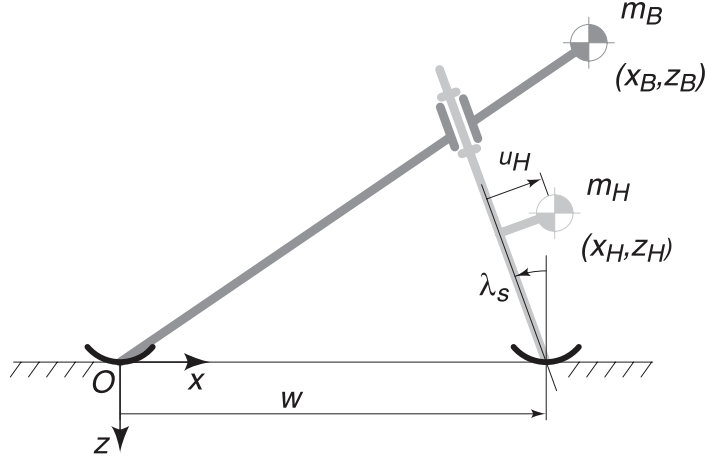}
    \end{center}
    \caption{The two-mass-skate (TMS) bicycle model \cite{TMS2011}.}
    \label{figTMS}
    \end{figure}

The TMS  model is sketched in Fig.~\ref{figTMS}. It depends on 9 dimensional parameters:
$$
w, \quad v, \quad \lambda_s,\quad m_B,\quad x_B, \quad z_B,\quad m_H,\quad x_H,\quad z_H
$$
that represent, respectively, the wheel base, velocity of the bicycle, steer axis tilt, rear frame assembly ($B$) mass, horizontal and vertical coordinates
of the rear frame assembly center of mass, front fork and handlebar assembly ($H$) mass, and horizontal and vertical coordinates
of the front fork and handlebar assembly center of mass \cite{TMS2011}.

We wish to study stability of the TMS bicycle that is moving along a straight trajectory with the constant velocity and remaining in a straight vertical position. In order to simplify the analysis it is convenient to choose the wheelbase, $w$, as a unit of length and introduce the dimensionless time $\tau =t\sqrt{\frac{g}{w}}$ and 7 dimensionless parameters
$$
\mathrm{Fr}=\frac{v}{\sqrt{g w}},~~ \mu=\frac{m_H}{m_B},~~ \chi_B=\frac{x_B}{w}, ~~ \chi_H=\frac{x_H}{w}, ~~
\zeta_B=\frac{z_B}{w}, ~~ \zeta_H=\frac{z_H}{w}, ~~ \lambda_s,
$$
where $g$ is the gravity acceleration, $\mathrm{Fr}$ the Froude number and $\mu$ the mass ratio. Notice that $\zeta_B\le 0$ and $\zeta_H\le 0$
due to choice of the system of coordinates, Fig.~\ref{figTMS}.

It has been shown in \cite{Ruina2007,TMS2011} that small deviations from the straight vertical equilibrium of the TMS bicycle are described by the leaning angle, $\phi$, of the
frame and the steering angle, $\delta$, of the front wheel/skate. These angles are governed by the two coupled linear equations
\be{ur}
\textbf{M} \ddot{\textbf{q}} + \textbf{V} \dot{\textbf{q}} + \textbf{P}{\textbf{q}} =0,\quad {\bf q}=(\phi, \delta)^T,
\ee
where dot denotes differentiation with respect to dimensionless time, $\tau$, and the matrices of mass, $\bf M$, velocity-dependent forces, $\bf V$, and positional forces, $\bf P$, are
\ba{matrices}
&{\bf M}=\left(
  \begin{array}{cc}
    \mu\zeta_H^2+\zeta_B^2 & -\mu\zeta_H\nu_H \\
    -\mu\zeta_H\nu_H & \mu \nu_H^2 \\
  \end{array}
\right),\quad
{\bf V}=\left(
  \begin{array}{cc}
    0 & -\mu\chi_H\zeta_H-\chi_B\zeta_B \\
    0 & \mu\chi_H  \nu_H \\
  \end{array}
\right){\rm Fr}\cos\lambda_s,&\nn\\
&{\bf P}=\left(
          \begin{array}{cc}
            \mu\zeta_H+\zeta_B & -{\rm Fr}^2\cos\lambda_s(\mu\zeta_H+\zeta_B)-\mu \nu_H \\
            -\mu \nu_H & \mu({\rm Fr}^2\cos\lambda_s-\sin\lambda_s)\nu_H \\
          \end{array}
        \right)&
\ea
with $\nu_H=\frac{u_H}{w}=(\chi_H -1)\cos\lambda_s -\zeta_H\sin\lambda_s$, see Fig.~\ref{figTMS}.

\subsection{Asymptotic stability of the TMS bike and the critical Froude number for the weaving motion}

The TMS model \rf{ur}, \rf{matrices} is nonconservative, containing dissipative, gyroscopic, potential and circulatory (curl, \cite{BS2016}) forces.
Assuming the solution $\sim \exp (s \tau)$ we write the characteristic polynomial of the TMS bicycle model:
\be{chapo}
p(s)=a_0 s^4+a_1 s^3 + a_2 s^2 + a_3 s +a_4,
\ee
with the coefficients
\ba{copo}
a_0 &=& -(\zeta_H\tan\lambda_s-\chi_H+1)\zeta_B^2,\nn\\
a_1 &=& \mathrm{Fr}(\zeta_B\chi_H-\zeta_H\chi_B)\zeta_B,\nn\\
a_2 &=& \mathrm{Fr}^2(\zeta_B-\zeta_H)\zeta_B- \zeta_B(\zeta_B+\zeta_H)\tan\lambda_s-(\chi_H-1)(\mu\zeta_H-\zeta_B),\nn\\
a_3 &=& -\mathrm{Fr}(\chi_B-\chi_H)\zeta_B, \nn\\
a_4 &=& -\zeta_B\tan\lambda_s-\mu(\chi_H-1).
\ea

Applying the Lienard-Chipart version of the Routh-Hurwitz criterion \cite{K2013b,K2018} to the polynomial \rf{chapo} yields
for $\tan \lambda_s >0$ the following necessary and sufficient conditions for the asymptotic stability of the TMS bicycle
\ba{lscf}
\chi_H&>&1+\zeta_H\tan\lambda_s,\nn\\
\chi_H&<&   1-\frac{\zeta_B}{\mu}\tan\lambda_s,\nn\\
\chi_H&<&\chi_B,\nn\\
\zeta_H&>&\zeta_B,\nn\\
{\mathrm{Fr}}&>&{\mathrm{Fr}}_c>0,
\ea
where the critical Froude number at the stability boundary is given by the expression
\be{flus}
{\mathrm{Fr}}_c^2=\frac{\zeta_B-\zeta_H}{\chi_B-\chi_H}\frac{\chi_B\chi_H}{\zeta_B\chi_H-\zeta_H\chi_B}\tan\lambda_s
+\frac{\chi_H-1}{\chi_B-\chi_H}\frac{\chi_H}{\zeta_B}\mu-\frac{\chi_H-1}{\zeta_B\chi_H-\zeta_H\chi_B}\chi_B.
\ee
At $0\le {\mathrm{Fr}}<{\mathrm{Fr}}_c$ the bicycle is unstable while at ${\rm Fr}>{\rm Fr}_c$ it is asymptotically stable. The critical value ${\rm Fr}_c$ is on the boundary between the domains of the asymptotic stability and
dynamic instability (\textit{weaving motion}, \cite{Ruina2007,Sharp2008,TMS2011}). Notice that in the recent work \cite{GAS2018} a comprehensive analysis of the Lienard -Chipart conditions for the TMS-bicycle reduced self-stable designs to just two classes corresponding to either positive or negative angles $\lambda_s$ and excluded backward stability for the TMS model. In the following we will limit our analysis to the $(\lambda_s> 0)$-class of the self-stable TMS bikes.

For instance, for the wheel base $w=1m$ the design proposed in \cite{TMS2011} is determined by
\ba{bench}
\lambda_s=\frac{5\pi}{180}rad,~~ m_H=1kg,~~ m_B=10kg,~~
x_B=1.2m,~~ x_H=1.02m,~~ z_B=-0.4m,~~ z_H=-0.2m.
\ea
With \rf{bench} we find from \rf{flus} the critical Froude number and the corresponding critical velocity
\be{fr1}
\mathrm{Fr}_c=0.9070641497,~~v_c=2.841008324m/s
\ee
that reproduce the original result obtained numerically in \cite{TMS2011}.

\subsection{Minimizing the spectral abscissa of general TMS bikes}

The criterion \rf{lscf} guarantees asymptotic stability of the bicycle at ${\rm Fr}>{\rm Fr}_c$. However,
the character of time dependence of the steering and leaning angles could be different in different points within the domain of asymptotic stability. Indeed, complex eigenvalues with negative real parts correspond to exponentially decaying oscillatory motions whereas negative real eigenvalues yield exponential decay of perturbations without oscillations. If all the eigenvalues of the system are real and negative, the system is called {\em heavily damped} \cite{BL1992,V2011}. If we wish that the deviations from the straight vertical position of the heavily damped TMS bike riding along a straight line also quickly die out, we need to maximize the decay rates of the deviations in the following sense.

The abscissa of the polynomial $p(s)$ is the maximal real part of its roots
$$
a(p)=\max \left\{{\rm Re}~ s:~ p(s)=0 \right\}.
$$
Minimization of the spectral abscissa over the coefficients of the polynomial provides a polynomial with the roots that have minimal possible real parts (maximal possible decay rates). In the case of the system of coupled oscillators of the form \rf{ur} it is known that the minimum of the spectral abscissa is $a_{min}=\omega_0=-\sqrt[4]{\frac{\det {\bf P}}{\det {\bf M}}}$ \cite{FL1999}. Knowing the coefficients of the characteristic polynomial \rf{copo} it is easy to find that for the TMS bicycle
\be{w0}
\omega_0 = -\sqrt[4]{\frac{1}{\zeta_B^2}\frac{\zeta_B\tan\lambda_s+\mu(\chi_H-1)}{\zeta_H\tan\lambda_s-(\chi_H-1)}}.
\ee
Remarkably, if $s=\omega_0$ is the minimum of the spectral abscissa, it is the 4-fold root of the fourth-order characteristic polynomial \rf{chapo} which is the quadruple negative real eigenvalue with the Jordan block of order 4 of the linear operator ${\bf M}s^2+{\bf V}s+{\bf P}$ \cite{FL1999,KO2013}.
In this case the polynomial \rf{chapo} takes the form
\be{4poly}
p(s)=(s-\omega_0)^4=s^4-4s^3\omega_0+6s^2\omega_0^2-4s\omega_0^3+\omega_0^4,\quad \omega_0^4=\frac{a_4}{a_0}=\frac{\det {\bf P}}{\det {\bf M}}.
\ee
Comparing \rf{chapo} and \rf{4poly} we require that
\ba{a1a3}
a_1&=&\mathrm{Fr}(\zeta_B\chi_H-\zeta_H\chi_B)\zeta_B=-4\omega_0a_0,\nn\\
a_3&=& -\mathrm{Fr}(\chi_B-\chi_H)\zeta_B=-4\omega_0^3a_0.\nn
\ea
Dividing the first equation by the second one, we get the relation
$$
\frac{\zeta_B\chi_H-\zeta_H\chi_B}{\chi_B-\chi_H}=\frac{-1}{\omega_0^2}
$$
that we resolve with respect to $\chi_B$ to obtain the following {\em design constraint} (or {\em scaling law})
\be{chib}
\chi_B = \frac{\omega_0^2\zeta_B-1}{\omega_0^2\zeta_H-1} \chi_H.
\ee
Another constraint follows from the requirement $a_2=6\omega_0^2a_0$:
\be{a2}
 \mathrm{Fr}^2(\zeta_B-\zeta_H)+(6\omega_0^2\zeta_H\zeta_B- \zeta_B-\zeta_H)\tan\lambda_s=\zeta_B^{-1}(\chi_H-1)(6\omega_0^2\zeta_B^2+\mu\zeta_H-\zeta_B).
\ee

           \begin{figure}
    \begin{center}
    \includegraphics[angle=0, width=0.45\textwidth]{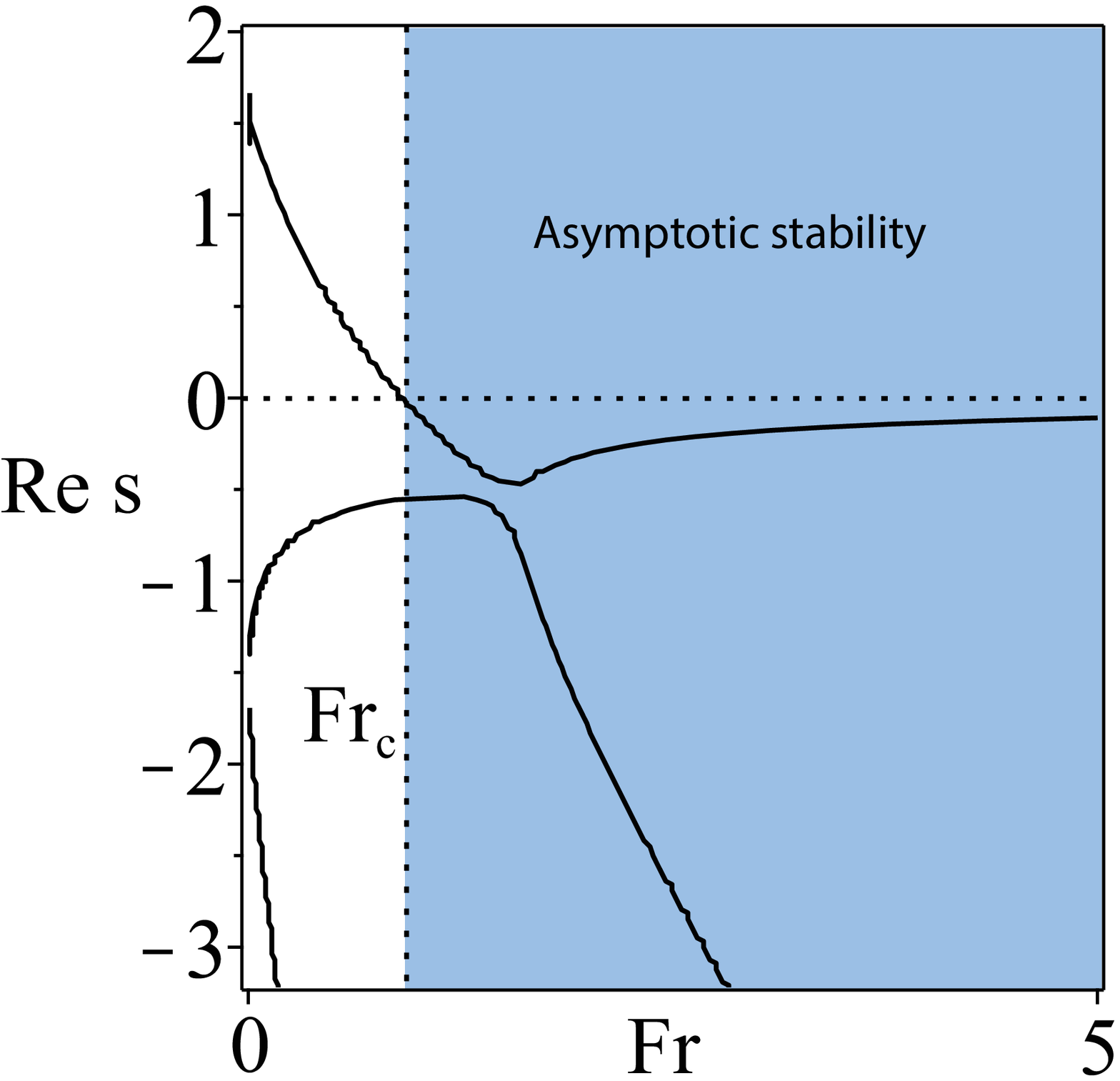}
    \includegraphics[angle=0, width=0.45\textwidth]{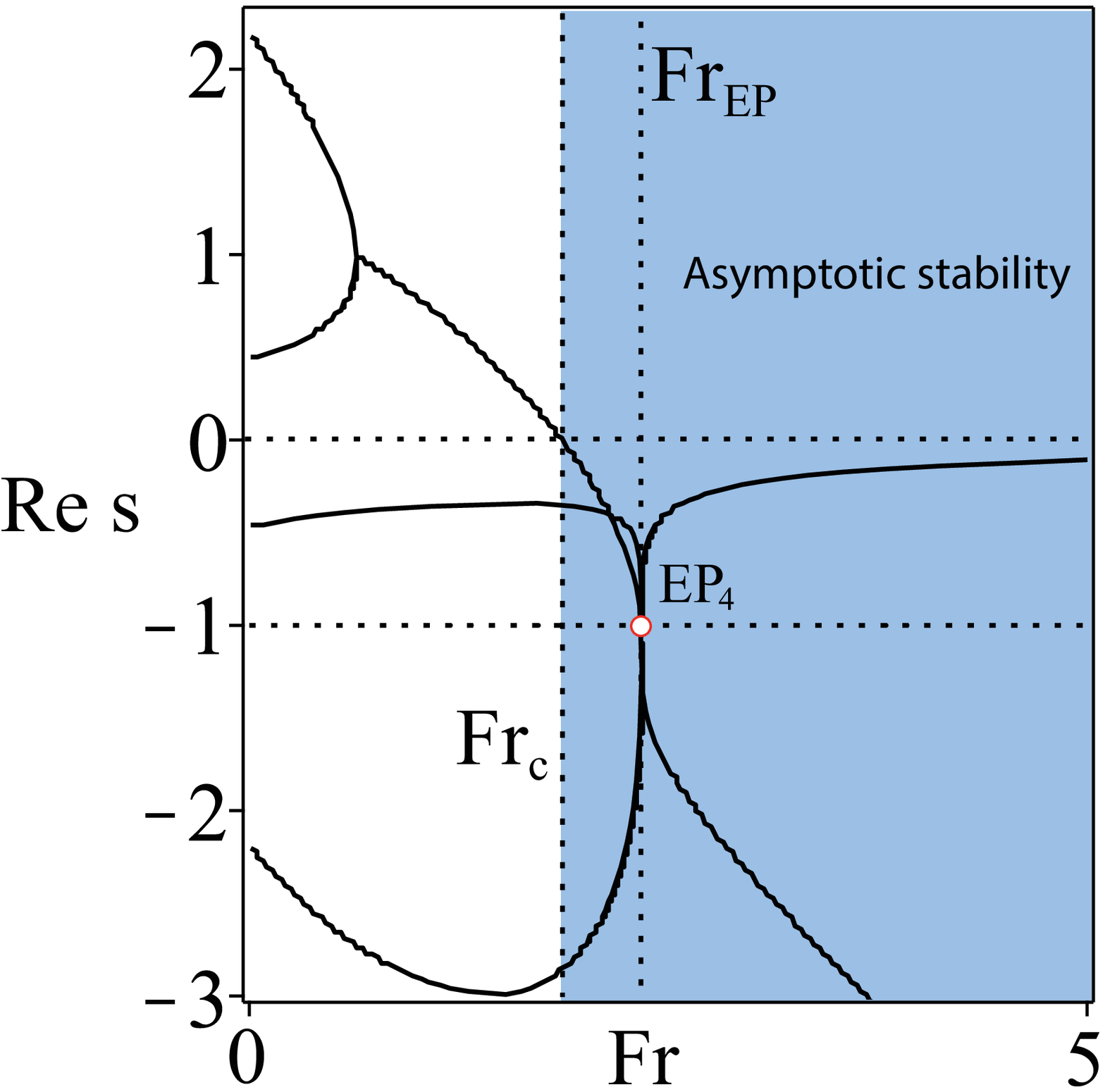}
    \end{center}
    \caption{(Left) The growth rates for the benchmark TMS bicycle \rf{bench}. (Right) The growth rates of the optimized TMS bicycle
    with $\zeta_B=-0.4$, $\zeta_H=-0.2$, $\chi_B=1.19$, $\chi_H=1.02$, $\mu = 20.84626701$, and $\lambda_s = 0.8514403685$ show that the spectral abscissa attains its minimal value $a_{min}=-1$ at ${\rm Fr}_{\rm EP}=2.337214017$ at the real exceptional point of order 4, ${\rm EP}_4$.}
    \label{fig0}
    \end{figure}

Let us optimize stability of the benchmark \rf{bench}. Set, for example, $\omega_0=-1$. Then, taking from the benchmark \rf{bench} the parameters $\zeta_B=-0.4$, $\zeta_H=-0.2$, and $\chi_H=1.02$ we find from Eq.~\rf{chib} that $\chi_B=1.19$. With these values the constraint \rf{a2} is
\be{e1}
-0.432\tan\lambda_s-0.0272+0.08{\rm Fr}^2+0.004\mu=0,
\ee
the relation \rf{w0} yields
\be{e2}
0.368\tan\lambda_s-0.02\mu-0.0032=0,
\ee
and the characteristic polynomial evaluated at $s=-1$ results in the equation
\be{e3}
0.192\tan\lambda_s-0.0048-0.136{\rm Fr}+0.08{\rm Fr}^2-0.016\mu=0.
\ee

The system \rf{e1}--\rf{e3} has a unique solution with the mass ratio $\mu>0$:
$$
{\rm Fr} = 2.337214017,\quad  \mu = 20.84626701, \quad \lambda_s = 0.8514403685.
$$
This means that the optimized TMS bicycle attains the minimum of the spectral abscissa
at ${\rm Fr}_{EP} = 2.337214017$ where all four eigenvalues merge into a quadruple negative real eigenvalue $s=-1$ with the Jordan block,
Fig.~\ref{fig0}(right). This eigenvalue we call a {\em real exceptional point of order 4} and denote as ${\rm EP}_4$. For comparison
we show in Fig.~\ref{fig0}(left) the growth rates of a generic benchmark TMS bicycle \rf{bench}.

Why the localization of the real ${\rm EP}_4$ is important? In \cite{KO2013} it was shown that this exceptional point is a Swallowtail singularity
on the boundary of the domain of heavy damping inside the domain of asymptotic stability of a system with two degrees of freedom. Furthermore, the minimum of the spectral abscissa occurs exactly at the Swallowtail degeneracy. In \cite{KO2013} it was shown that the ${\rm EP}_4$ `organizes'
asymptotic stability and its knowledge helps to localize other exceptional points governing stability exchange between modes. Below we demonstrate this explicitly for the TMS bikes with $\chi_H=1$.

\subsection{Self-stable and heavily damped TMS bikes with $\chi_H=1$}

\subsubsection{The critical Froude number and its minimum}

Why $\chi_H=1$? First, both the benchmarks reported in \cite{TMS2011} and their realizations have $\chi_H\approx 1$.
Second, this choice leads to a dramatic simplification without affecting generality of our consideration.
Indeed, the expression \rf{flus} for the critical Froude number evaluated at $\chi_H=1$ reduces to
\be{Frc1}
{\rm Fr}_c^2=\frac{\zeta_B-\zeta_H}{\zeta_B-\chi_B\zeta_H}\frac{\chi_B}{\chi_B-1}\tan\lambda_s.
\ee
Choosing $\chi_H=1$ automatically makes ${\rm Fr}_c$ and the stability conditions \rf{lscf} independent on the mass ratio $\mu$.
Additionally, the criteria \rf{lscf} imply $\chi_B>1$ and $|\zeta_B|>|\zeta_H|$.

Therefore, choosing $\chi_H= 1$ reduces the dimension of the parameter space from 7 to 5. The ($\chi_H= 1$)--bike depends just on
$\rm Fr$, $\chi_B$, $\zeta_H$, $\zeta_B$, and $\lambda_s$.

Given $\zeta_H$, $\zeta_B$, and $\lambda_s$ let us find the minimum of the critical Froude number \rf{Frc1} as a function of $\chi_B$.
It is easy to see that the minimum is attained at
\be{chibm}
\chi_B = \sqrt{\frac{\zeta_B}{\zeta_H}}
\ee
and its value is equal to
\be{frm}
{\rm Fr}_{min}=\sqrt{\frac{\sqrt{|\zeta_B|}+\sqrt{|\zeta_H|}}{\sqrt{|\zeta_B|}-\sqrt{|\zeta_H|}}\tan\lambda_s}.
\ee


These results suggest that all the critical parameters for the ($\chi_H= 1$)--bike can be expressed in a similar elegant manner by means of $\zeta_H$, $\zeta_B$, and $\lambda_s$  only. Let us check these expectations calculating the location of the real exceptional point ${\rm EP}_4$ for the ($\chi_H= 1$)--bike.

\subsubsection{Exact location of the real exceptional point ${\rm EP}_4$}

Indeed, with $\chi_H=1$ the expression \rf{w0} for the real negative quadruple eigenvalue at ${\rm EP}_4$
yields
\be{w01}
\omega_0 = -\sqrt[4]{\frac{1}{\zeta_B\zeta_H}}.
\ee
The design constraint \rf{chib} reduces to the scaling law
\be{chib1}
\chi_B =\sqrt{\frac{\zeta_B}{\zeta_H}}
\ee
which is nothing else but the minimizer \rf{chibm} of the critical Froude number !
Solving simultaneously the equation \rf{a2} and the equation $p(\omega_0)=0$ we find explicitly
the second design constraint that determines $\tan \lambda_s$ at ${\rm EP}_4$:
\be{tls}
\tan \lambda_s=\frac{\omega_0^2(\zeta_B-\zeta_H)}
{16\zeta_H}\frac{(\zeta_B+\zeta_H)\omega_0^2-6}{(\zeta_B+\zeta_H)\omega_0^2-2}.
\ee
Finally, from the same system of equations we find that the Froude number at ${\rm EP}_4$, ${\rm Fr}_{{\rm EP}_4}$, is a root of the quadratic equation
\be{frep4}
\left(\omega_0^2\zeta_B-1\right){\rm Fr}_{{\rm EP}_4}^2+2\omega_0^3\zeta_B{\rm Fr}_{{\rm EP}_4}-(\omega_0^2\zeta_B+1)\tan\lambda_s=0,
\ee
where $\omega_0$ is given by equation \rf{w01} and $\tan\lambda_s$ by equation \rf{tls}.

\begin{table}
\begin{tabular}{ccccccccc}
\hline
\small
\textbf{Bike} & \textbf{$\chi_H$}	& \textbf{$\chi_B$}	& \textbf{$\zeta_H$} & \textbf{$\zeta_B$}& $\omega_0$ & \textbf{$\lambda_s$} (rad.) & \textbf{${\rm Fr}_c$} & \textbf{${\rm Fr}_{\rm EP}$}  \\
\hline
\\[-2.5ex]
${\rm EP}_4$ & 1		& $\sqrt{2}$	& $-0.2$	& $-0.4$	& $-\frac{\sqrt{5}}{\sqrt[4]{2}}$ & $ \arctan\left(\frac{15}{4}-\frac{75}{32}\sqrt{2}\right)~~~~~~~~~$ & $\frac{\sqrt{30\sqrt{2}+120}}{8}$ & $\frac{3\sqrt{110\sqrt{2}-120}}{8}$\\
$2{\rm EP}_2$& 1		& $\sqrt{2}$	& $-0.2$	& $-0.4$	& $-\frac{\sqrt{5}}{\sqrt[4]{2}}$  &$\arctan\left(\frac{15}{4}-\frac{75}{32}\sqrt{2}\right)-0.05$ & $\approx1.482682090$ & $\approx2.257421384$ \\
${\rm CEP}_2$& 1		& $\sqrt{2}$	& $-0.2$	& $-0.4$ &	$-\frac{\sqrt{5}}{\sqrt[4]{2}}$  & $\arctan\left(\frac{15}{4}-\frac{75}{32}\sqrt{2}\right)+0.80$ & $\approx3.934331969$ & $\approx4.103508160$\\
\hline
\label{tab1}
\end{tabular}
\caption{TMS bike designs with $\chi_H=1$}
\end{table}

Let us take $\zeta_H=-0.2$ and $\zeta_B=-0.4$ as in the benchmark \rf{bench}. Then \rf{chib1}, \rf{tls}, and \rf{frep4} localize the ${\rm EP}_4$ in the space of the parameters giving (Table~\ref{tab1})
$$
\chi_B=\sqrt{2}, \quad \tan \lambda_s=\frac{15}{4}-\frac{75}{32}\sqrt{2}, \quad {\rm Fr}_{{\rm EP}_4}=\frac{3\sqrt{110\sqrt{2}-120}}{8}\approx2.236317517.
$$

           \begin{figure}
    \begin{center}
    \includegraphics[angle=0, width=0.45\textwidth]{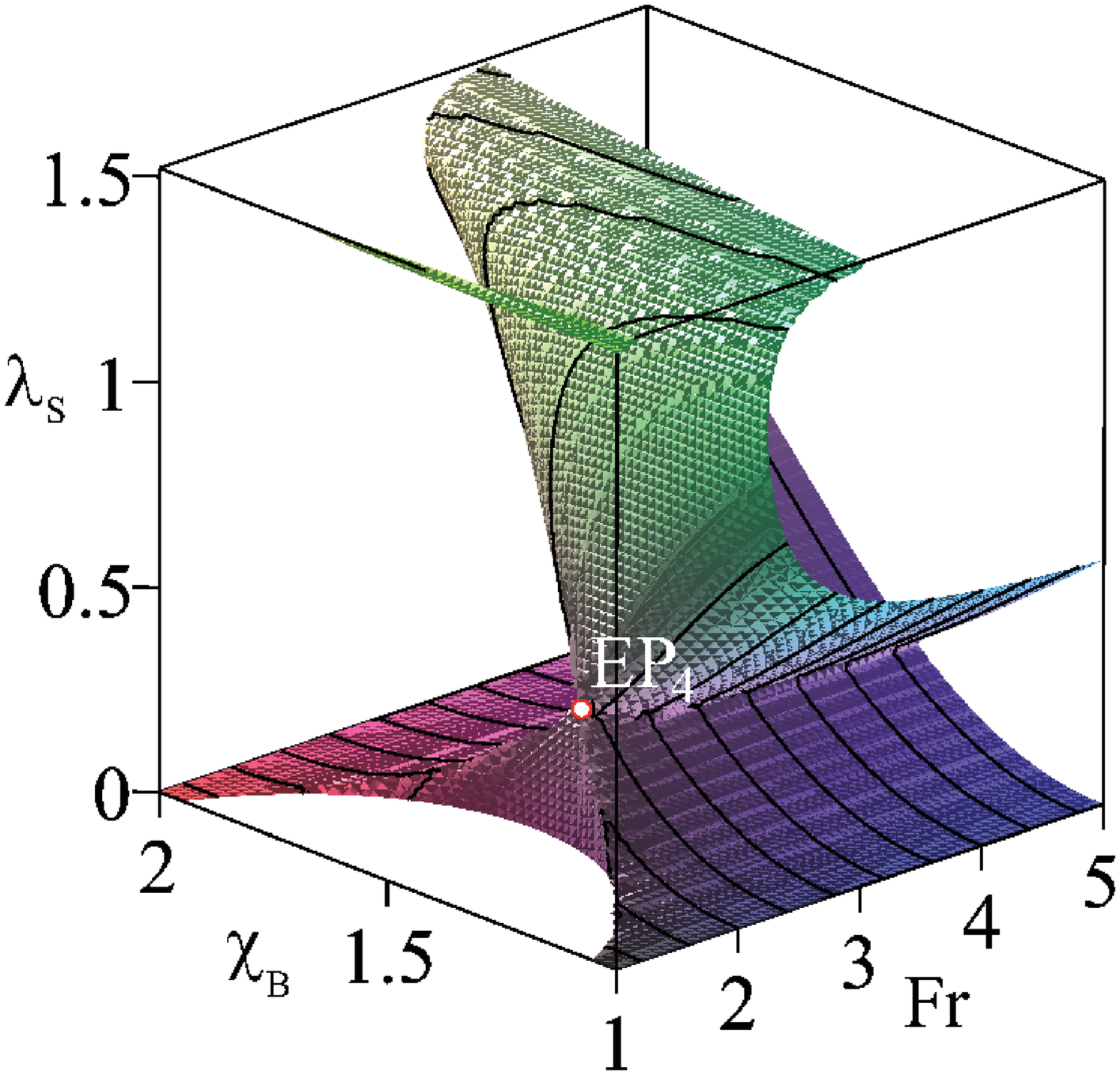}
    \includegraphics[angle=0, width=0.45\textwidth]{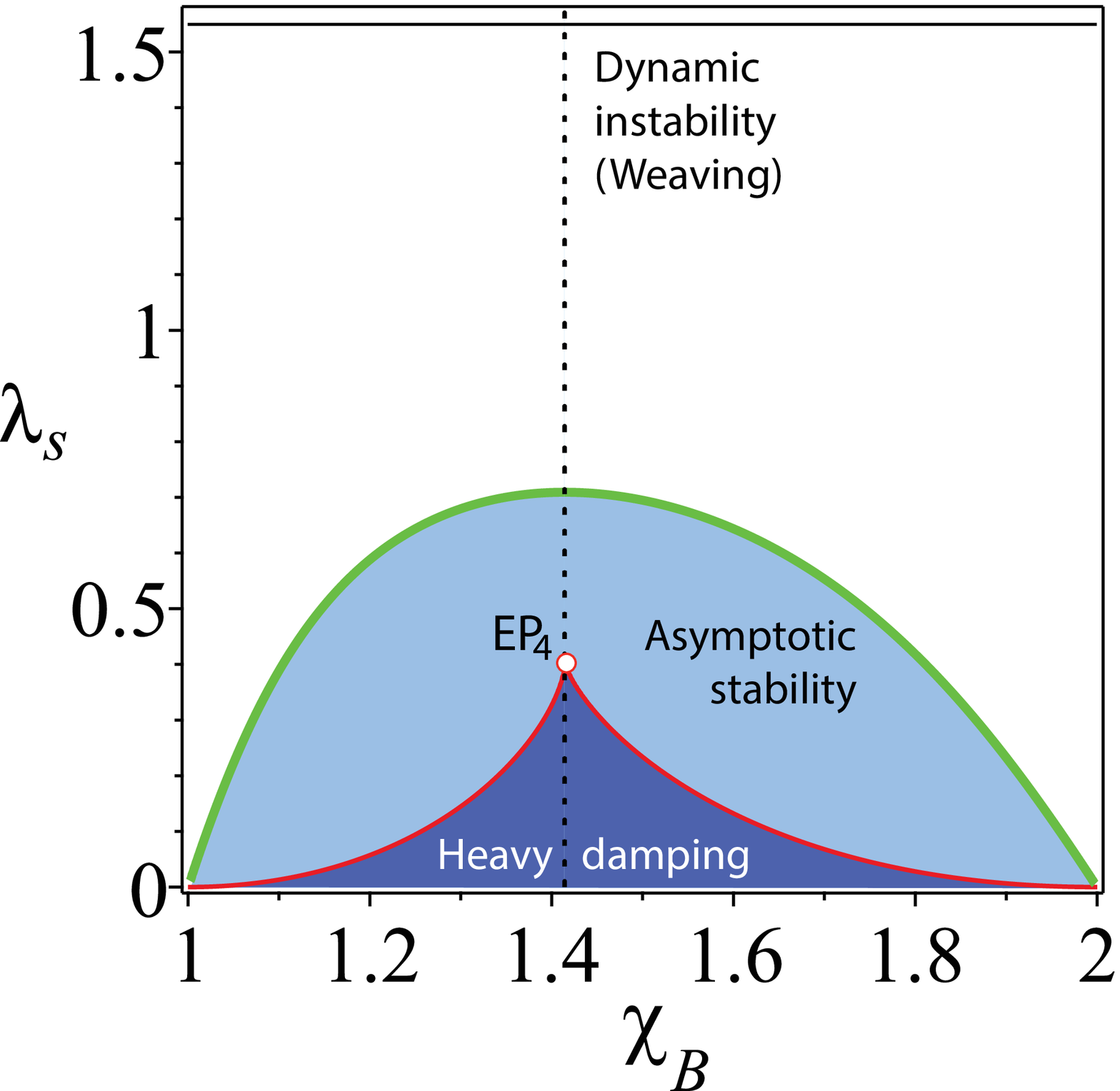}
    \includegraphics[angle=0, width=0.45\textwidth]{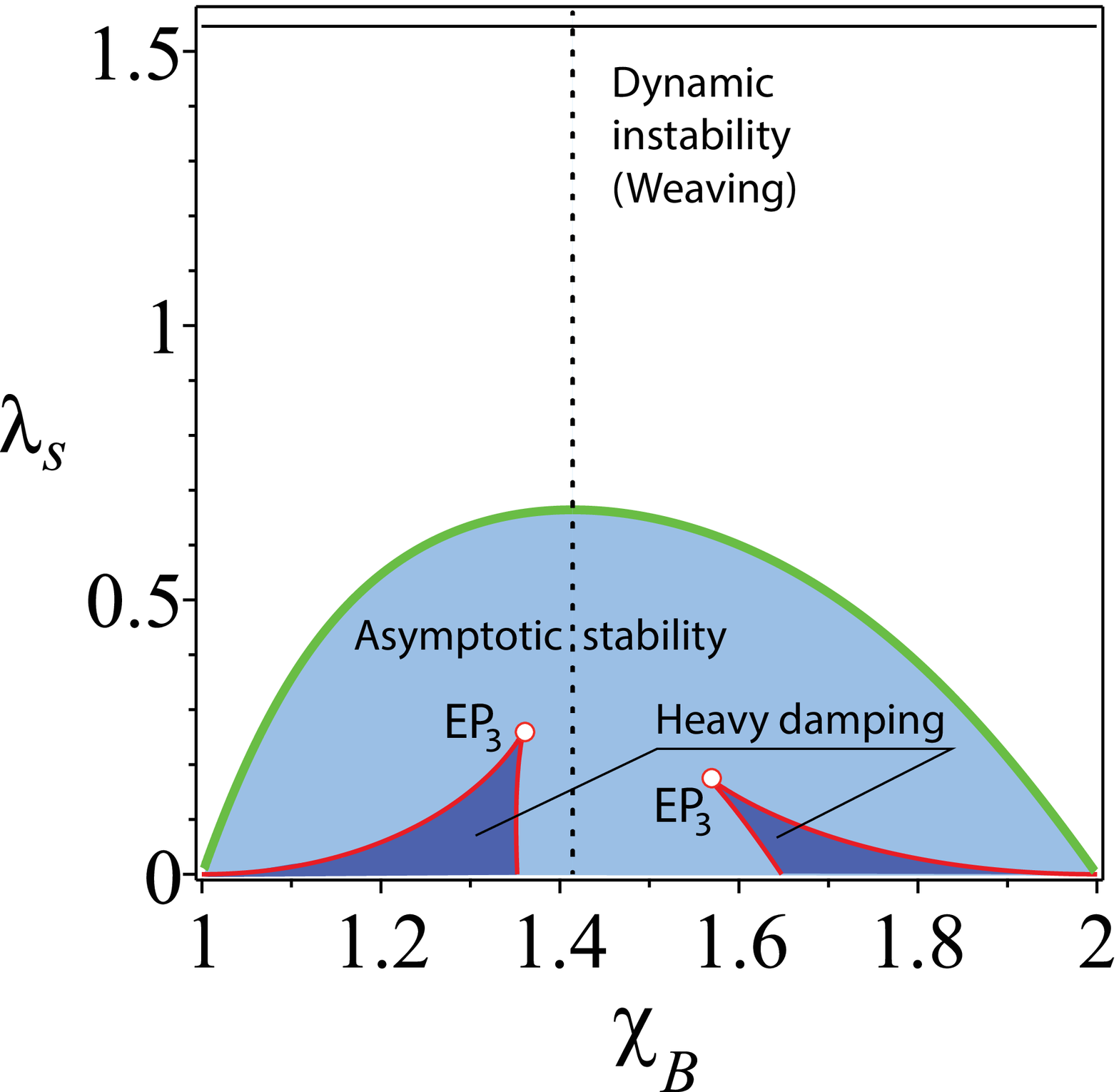}
    \includegraphics[angle=0, width=0.45\textwidth]{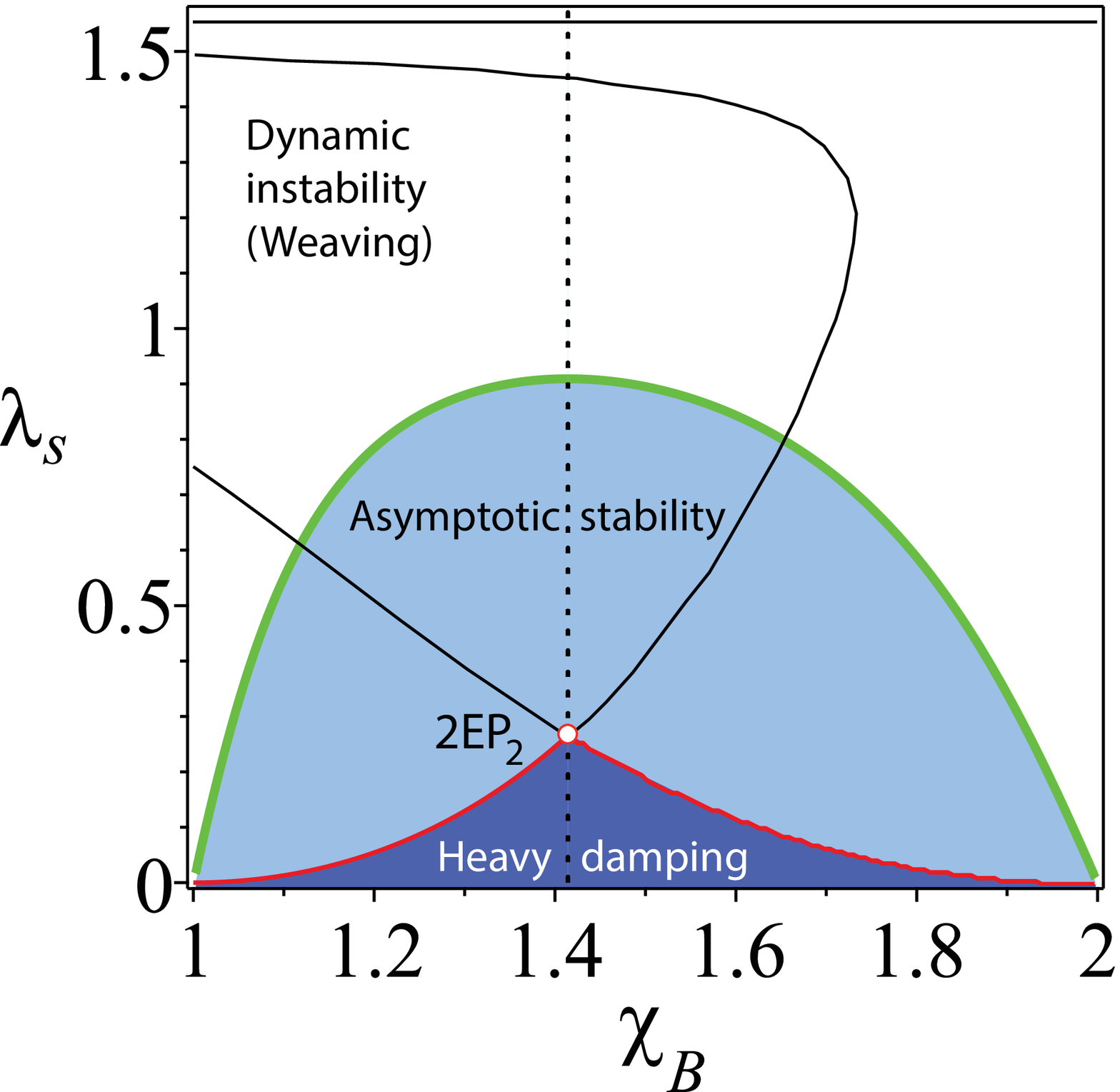}
    \end{center}
    \caption{(Upper left) The discriminant surface of the characteristic polynomial of the TMS-bike with $\chi_H=1$, $\zeta_H=-0.2$, and $\zeta_B=-0.4$ showing the Swallowtail singularity at ${\rm EP}_4$. The cross-section of the domain of asymptotic stability and the discriminant surface at (upper right) ${\rm Fr}={\rm Fr}_{{\rm EP}_4}=\frac{3\sqrt{110\sqrt{2}-120}}{8}$, (lower left) ${\rm Fr}={\rm Fr}_{{\rm EP}_4}-0.1$ and (lower right) ${\rm Fr}={\rm Fr}_{{\rm EP}_4}+0.5$. }
    \label{fig00}
    \end{figure}

\subsubsection{Discriminant surface and the EP-set}

The localized ${\rm EP}_4$ corresponds to a quadruple negative eigenvalue $s=\omega_0=-\frac{\sqrt{5}}{\sqrt[4]{2}}$. It is known that ${\rm EP}_4$ is the Swallowtail singular point on the discriminant surface of the fourth-order characteristic polynomial \cite{KO2013}. In Fig.~\ref{fig00} the discriminant surface is plotted in the $({\rm Fr},\chi_B,\lambda_s)$ --space for the TMS-bike with $\chi_H=1$, $\zeta_H=-0.2$, and $\zeta_B=-0.4$ showing the Swallowtail singular point with the position specified by the first line of the Table~\ref{tab1}. The discriminant surface has two cuspidal edges as well as the line of self-intersection branching from the ${\rm EP}_4$. These singularities belong to a boundary of a domain with the shape of a trihedral spire. This is the domain of heavy damping. In its inner points all the eigenvalues are real and negative \cite{KO2013}.

           \begin{figure}
    \begin{center}
    \includegraphics[angle=0, width=0.45\textwidth]{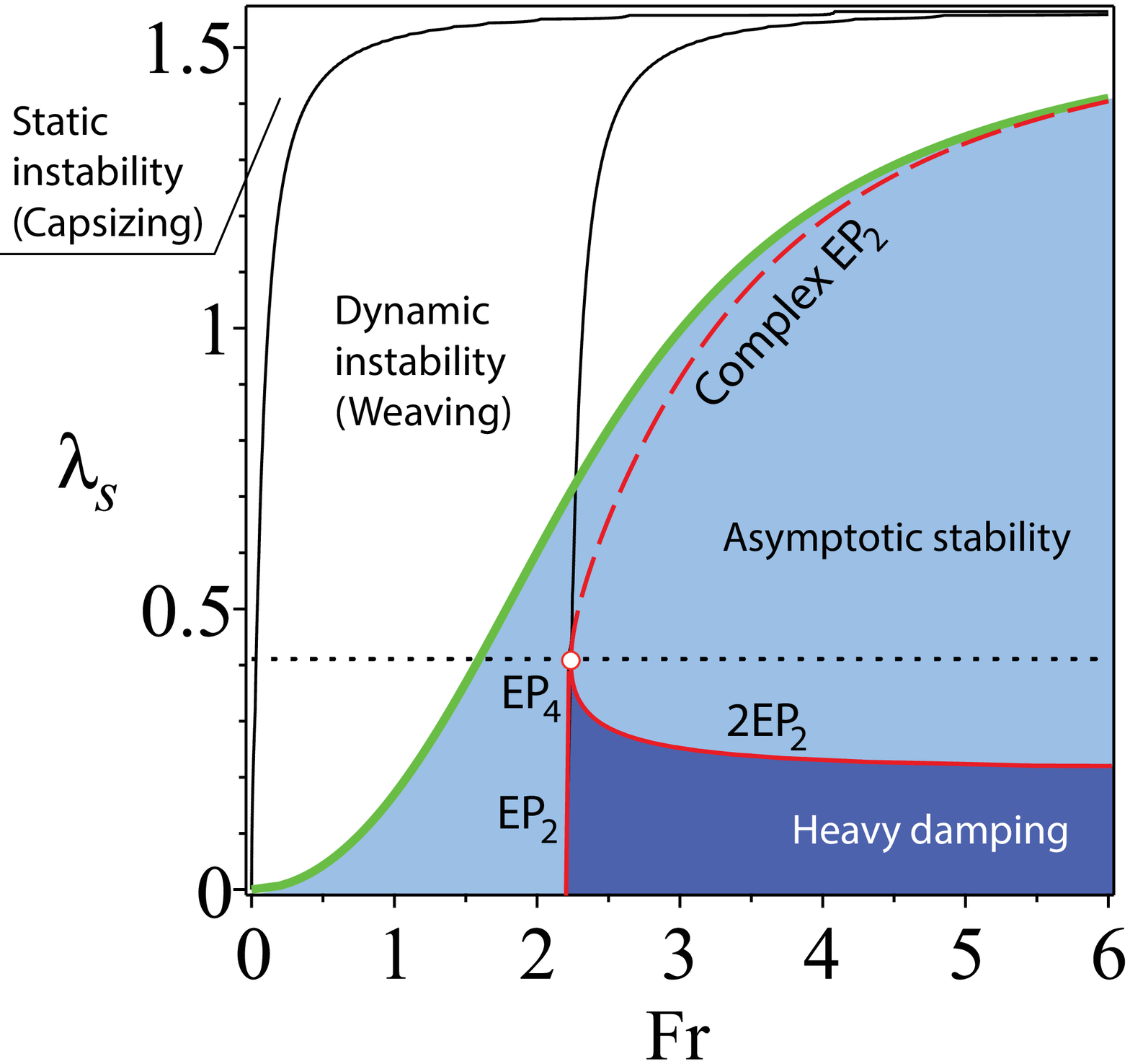}
    \includegraphics[angle=0, width=0.45\textwidth]{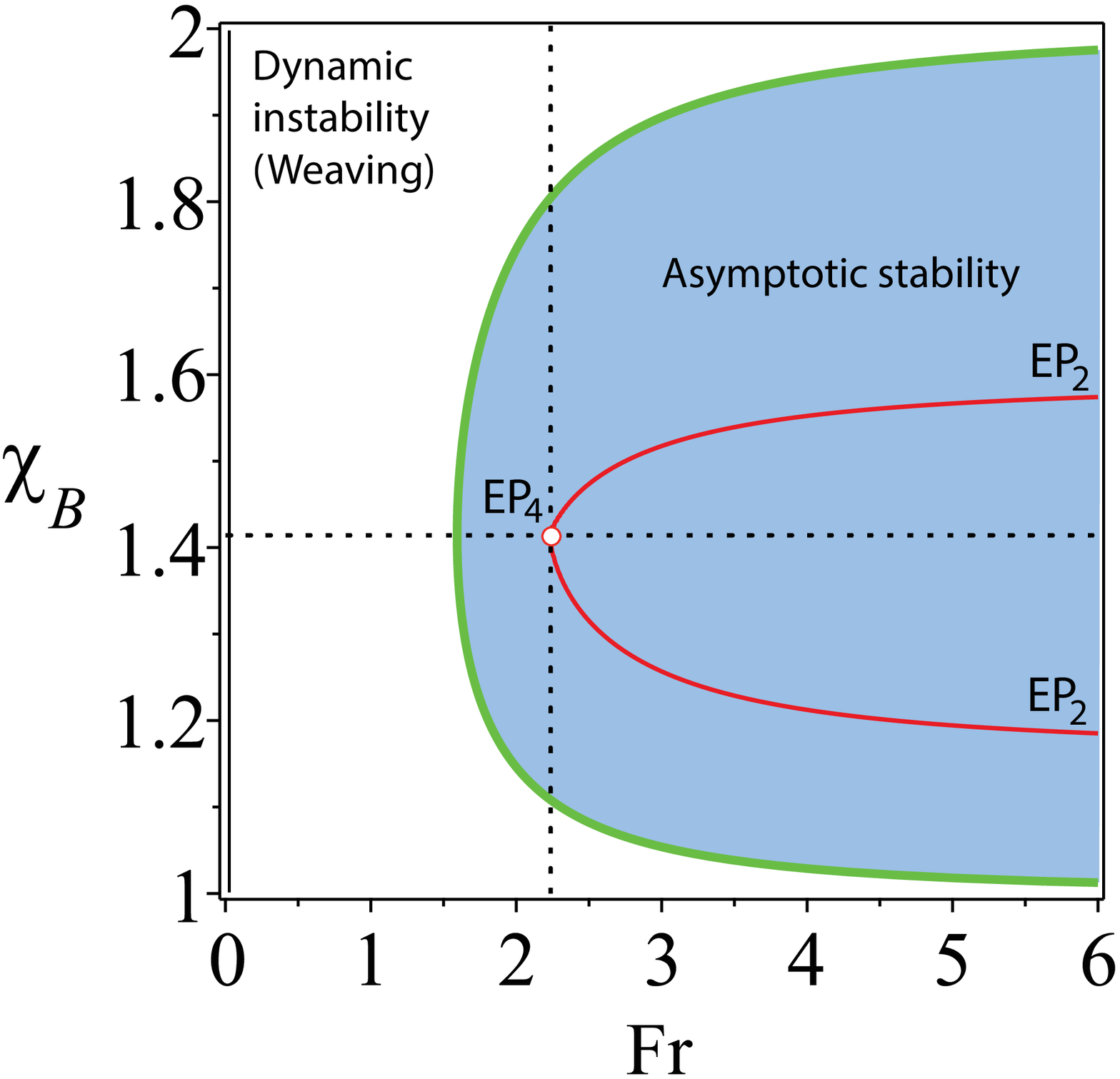}
    \end{center}
    \caption{$\chi_H=1$, $\zeta_H=-0.2$, and $\zeta_B=-0.4$.  (Left) For $\chi_B=\sqrt{2}$ the boundary between the domains of weaving and asymptotic stability in the $({\rm Fr},\lambda_s)$ - plane shown together with the domain of heavy damping that has a cuspidal point corresponding to a negative real eigenvalue $\omega_0 = -\sqrt[4]{\frac{25}{2}}$ with the Jordan block of order four (${\rm EP_4}$). The ${\rm EP_4}$ belongs to a curve \rf{epset} that corresponds to (dashed part) conjugate pairs of double complex eigenvalues with the Jordan block of order two (complex ${\rm EP_2}$) and (solid part) to couples of double real negative eigenvalues with the Jordan block of order two ($2{\rm EP_2}$). (Right) The same in the $({\rm Fr},\chi_B)$--plane at $\lambda_s^0= \arctan\left(\frac{15}{4}-\frac{75}{32}\sqrt{2}\right)$ rad. The domain of heavy damping degenerates into a singular point - the Swallowtail singularity.}
    \label{fig1}
    \end{figure}

We see that the line of self-intersection lies in the plane $\chi_B=\sqrt{\frac{\zeta_B}{\zeta_H}}$. Restricted to this plane (parameterized by $\rm Fr$ and $\lambda_s$) the discriminant of the characteristic polynomial \rf{chapo} simplifies and provides the following expression for the curve that contains the line of self-intersection of the discriminant surface
\be{epset}
{\rm Fr}=\frac{\omega_0^2\zeta_B-1}{\omega_0^2\zeta_B+1}\frac{2\tan\lambda_s}{\sqrt{
\omega_0^4\zeta_B+4\tan\lambda_s\frac{\omega_0^2\zeta_B-1}{\omega_0^2\zeta_B+1}}}.
\ee
In Fig.~\ref{fig1}(left) the curve \rf{epset} is plotted for $\chi_H=1$, $\zeta_H=-0.2$, $\zeta_B=-0.4$ and $\chi_B=\sqrt{2}$ in the $({\rm Fr},\lambda_s)$--plane. A point where this curve has a vertical tangent is the Swallowtail singularity or ${\rm EP}_4$. The part of the curve below the ${\rm EP}_4$ is a line of self-intersection of the discriminant surface corresponding to a pair of different negative double real eigenvalues with the Jordan block, i.e. to a couple of real exceptional points which we denote as $2{\rm EP}_2$.

The curve \rf{epset} continues, however, also above the ${\rm EP}_4$. This part shown by a dashed line in Fig.~\ref{fig1}(left) is the set corresponding to conjugate pairs of complex double eigenvalues with the Jordan block, or complex exceptional points that we denote as ${\rm CEP}_2$. Since the curve \rf{epset} is a location of three types of exceptional points we call it the {\em EP-set}. Notice that the codimension of the EP-set is 2 and by this reason its localization by numerical approaches is very complicated.

           \begin{figure}
    \begin{center}
    \includegraphics[angle=0, width=0.45\textwidth]{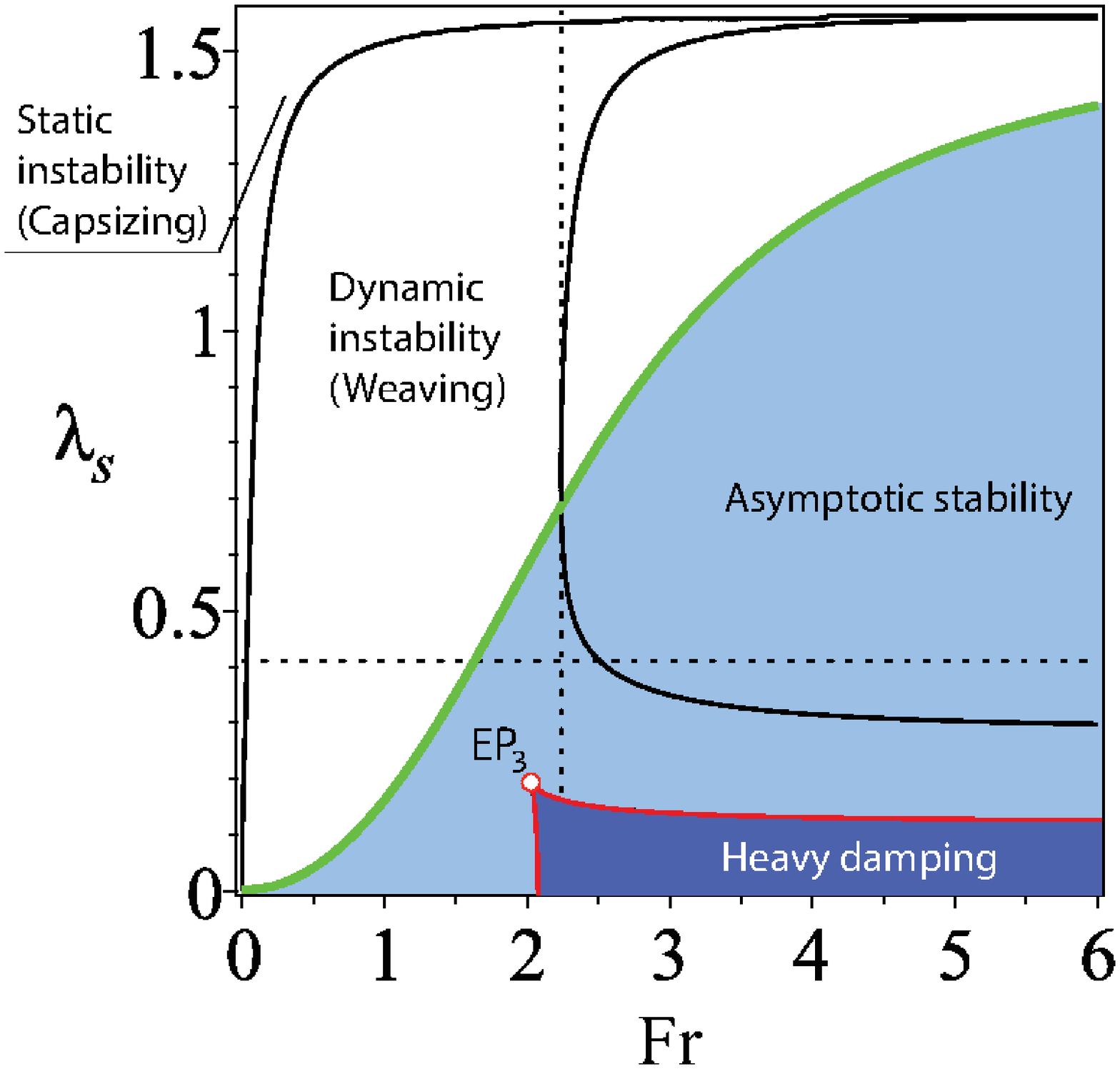}
    \includegraphics[angle=0, width=0.45\textwidth]{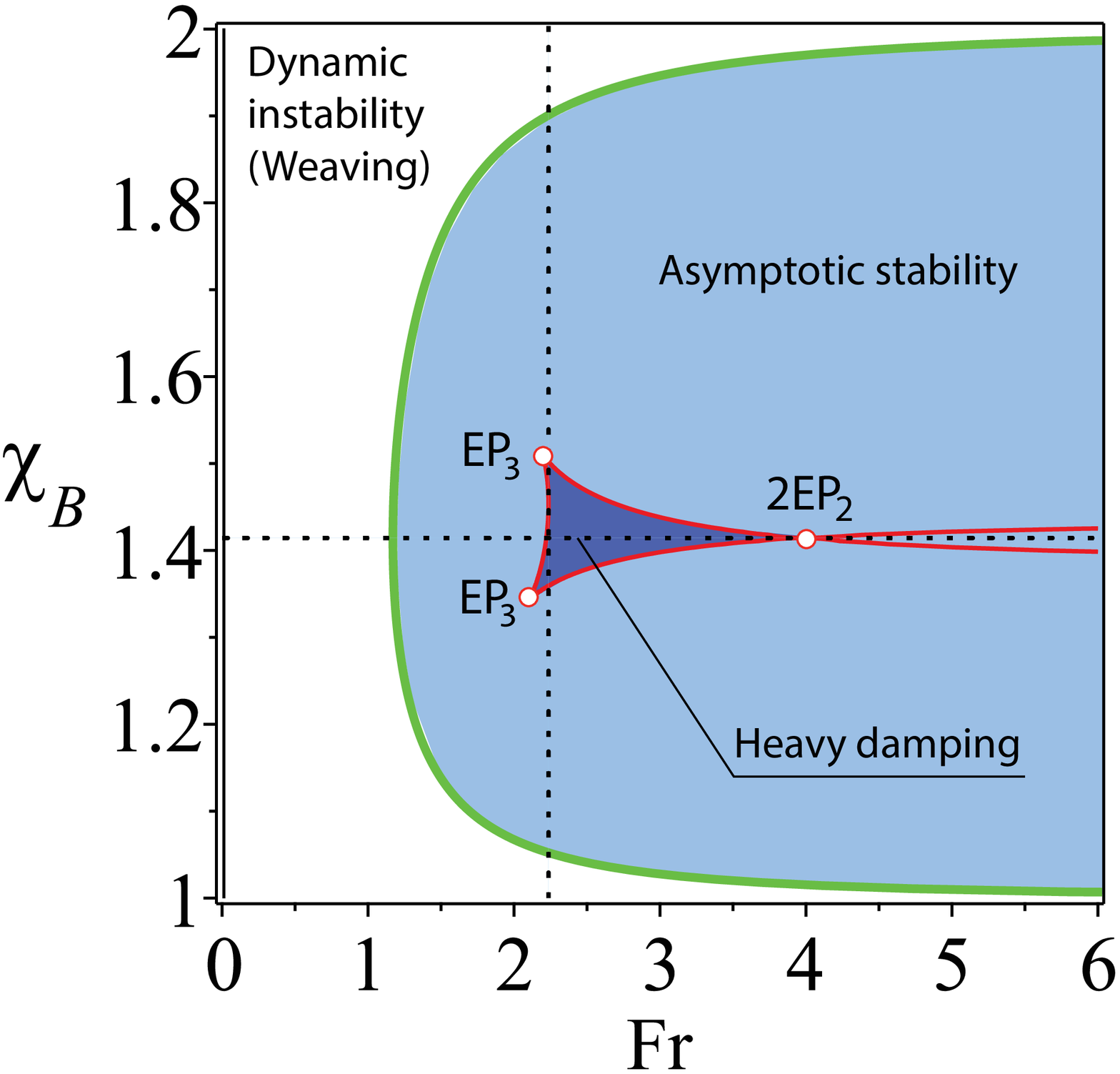}
    \end{center}
    \caption{$\chi_H=1$, $\zeta_H=-0.2$, and $\zeta_B=-0.4$.  (Left) For $\chi_B=\sqrt{2}-0.1$ the boundary between the domains of weaving and asymptotic stability in the $({\rm Fr},\lambda_s)$ - plane shown together with the domain of heavy damping that has a cusp corresponding to a negative real eigenvalue with the Jordan block of order three (${\rm EP_3}$). The ${\rm EP_3}$ belongs to the cuspidal edge of the swallowtail surface bounding the domain of heavy damping. (Right) The same in the $({\rm Fr},\chi_B)$-plane at $\lambda_s=\arctan\left(\frac{15}{4}-\frac{75}{32}\sqrt{2}\right)-0.18$ rad. Notice the cuspidal ${\rm EP_3}$-points and the self intersection at the $2{\rm EP_2}$ point on the boundary of the domain of heavy damping.}
    \label{fig2}
    \end{figure}

\subsubsection{Localization of the EP-set and stability optimization}

What the localization of the EP-set means for the stability of the TMS bike? Drawing the domain of asymptotic stability
together with the discriminant surface and the EP-set in the same plot, we see that the EP-set lies entirely
in the domain of asymptotic stability, Fig.~\ref{fig1}. The ${\rm 2EP}_2$ part of the EP-set bounds the domain of heavy damping in the plane $\chi_B=\sqrt{\frac{\zeta_B}{\zeta_H}}=\sqrt{2}$.

Look now at the cross-sections of the asymptotic stability domain and the discriminant surface in the $(\chi_B, \lambda_s)$--plane, Fig.~\ref{fig00}. Remarkably, the value $\chi_B=\sqrt{\frac{\zeta_B}{\zeta_H}}=\sqrt{2}$ is a maximizer of
the steer axis tilt $\lambda_s$ both at the onset of the weaving instability and at the boundary of the domain of heavy damping. In the latter case the maximum is always attained at a singular point in the EP-set: either at ${\rm EP}_4$ when ${\rm Fr}={\rm Fr}_{{\rm EP}_4}$ or at $2{\rm EP}_2$
when ${\rm Fr}>{\rm Fr}_{{\rm EP}_4}$. The absolute maximum of the steer axis tilt is attained exactly at ${\rm EP}_4$ which also minimizes the spectral abscissa. Taking into account that $\chi_B=\sqrt{\frac{\zeta_B}{\zeta_H}}=\sqrt{2}$ is a minimizer of the critical Froude number that is necessary for asymptotic stability, we conclude that the both of the design constraints, \rf{chib1} and \rf{tls}, play an important part in the self-stability phenomenon:

\textit{The most efficient self-stable TMS bikes are those that have better chance to operate in the heavy damping domain. In the case when $\chi_H=1$, these bikes should follow the scaling laws
\be{dp}\chi_B=\sqrt{\frac{\zeta_B}{\zeta_H}}\quad and \quad  0< \tan\lambda_s\le\frac{\omega_0^2(\zeta_B-\zeta_H)}
{16\zeta_H}\frac{(\zeta_B+\zeta_H)\omega_0^2-6}{(\zeta_B+\zeta_H)\omega_0^2-2}, \quad where \quad \omega_0=-\sqrt[4]{\frac{1}{\zeta_B\zeta_H}}.\ee}

Even in the case of an approximate scaling law $\chi_B \approx \sqrt{\frac{\zeta_B}{\zeta_H}}$ the domain of heavy damping is large enough, Fig.~\ref{fig2}, suggesting that the formulated principle produces sufficiently robust design of self-stable TMS bikes.

         \begin{figure}
    \begin{center}
    \includegraphics[angle=0, width=0.35\textwidth]{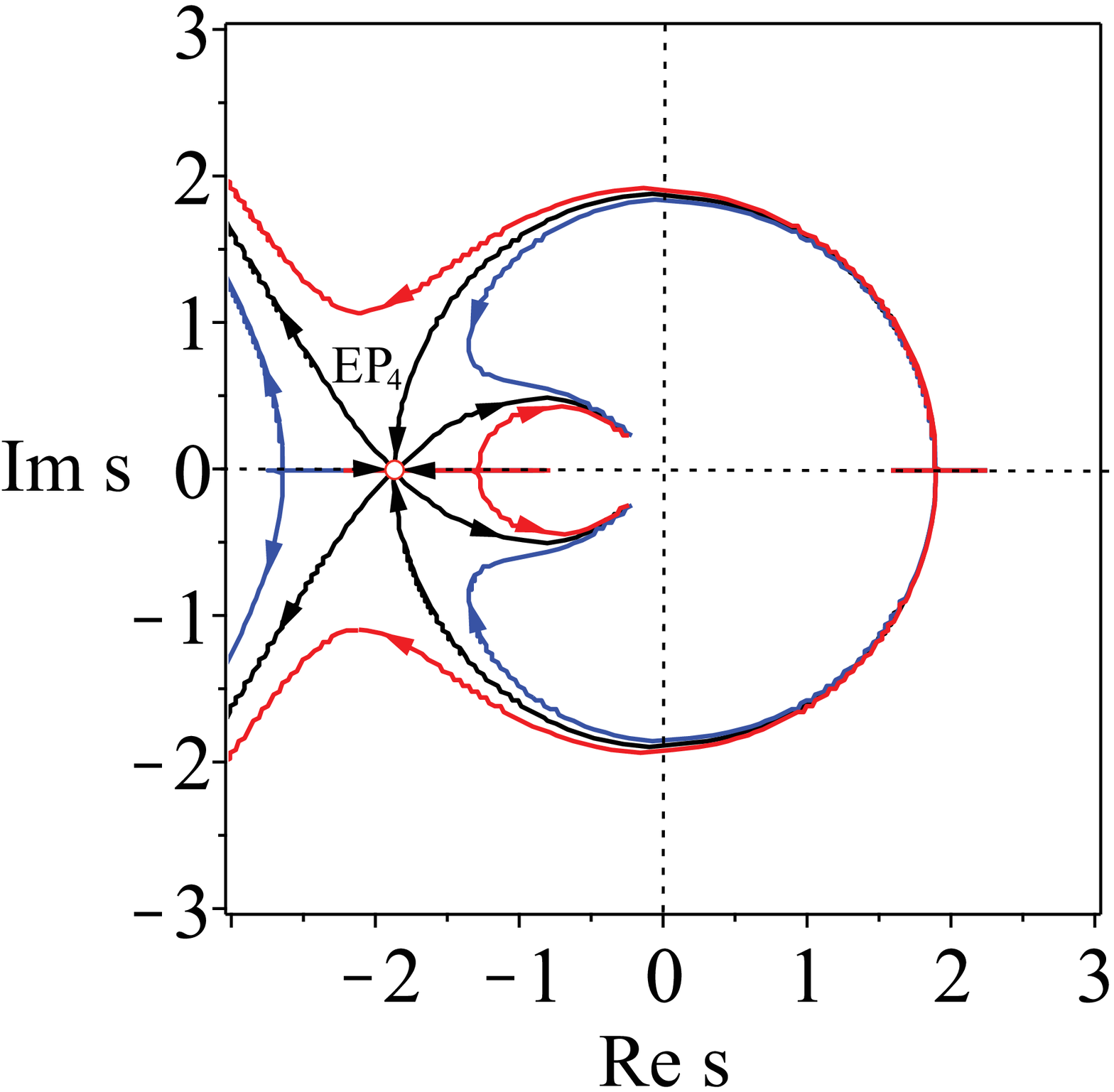}
    \includegraphics[angle=0, width=0.35\textwidth]{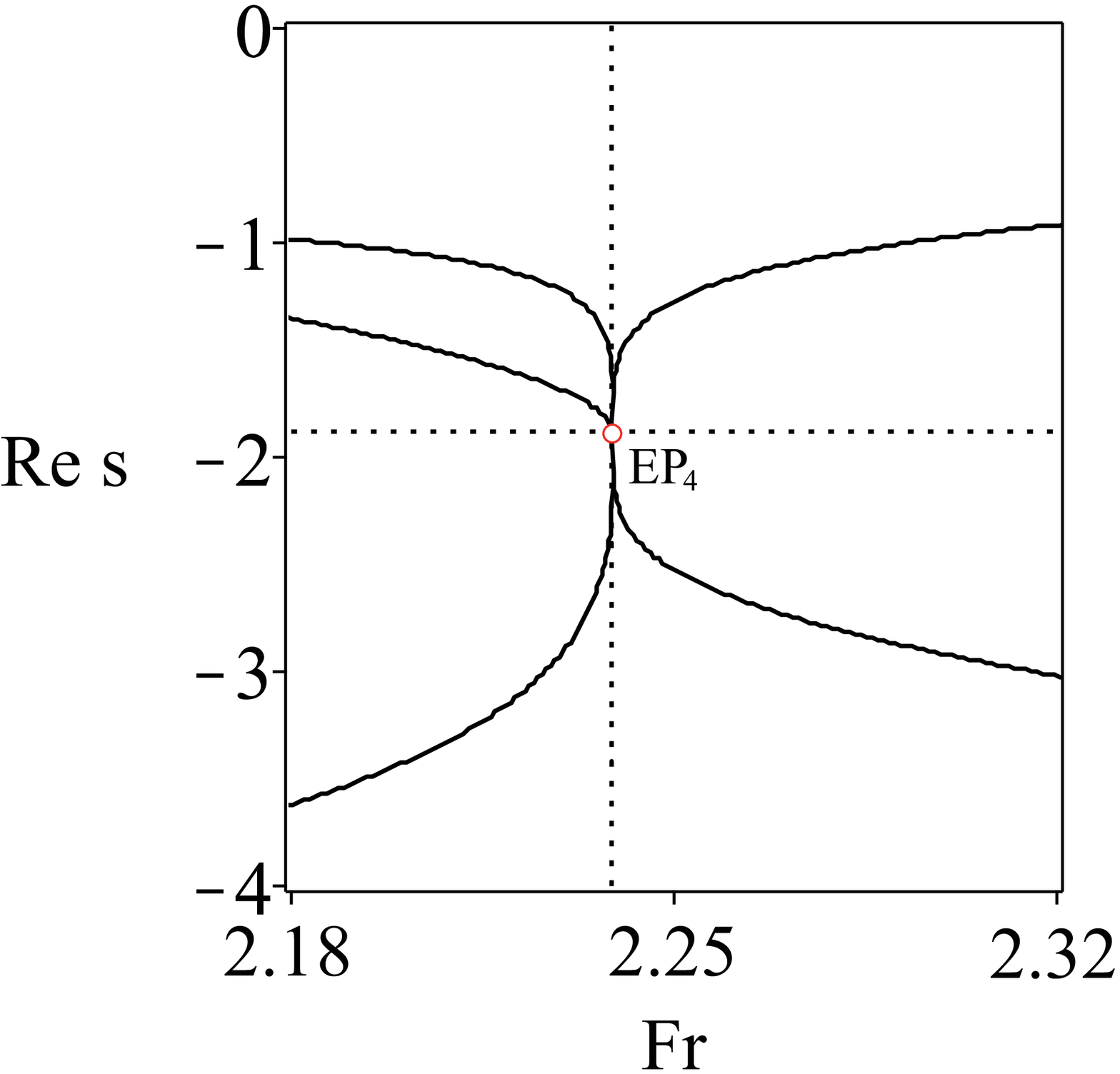}
    \includegraphics[angle=0, width=0.35\textwidth]{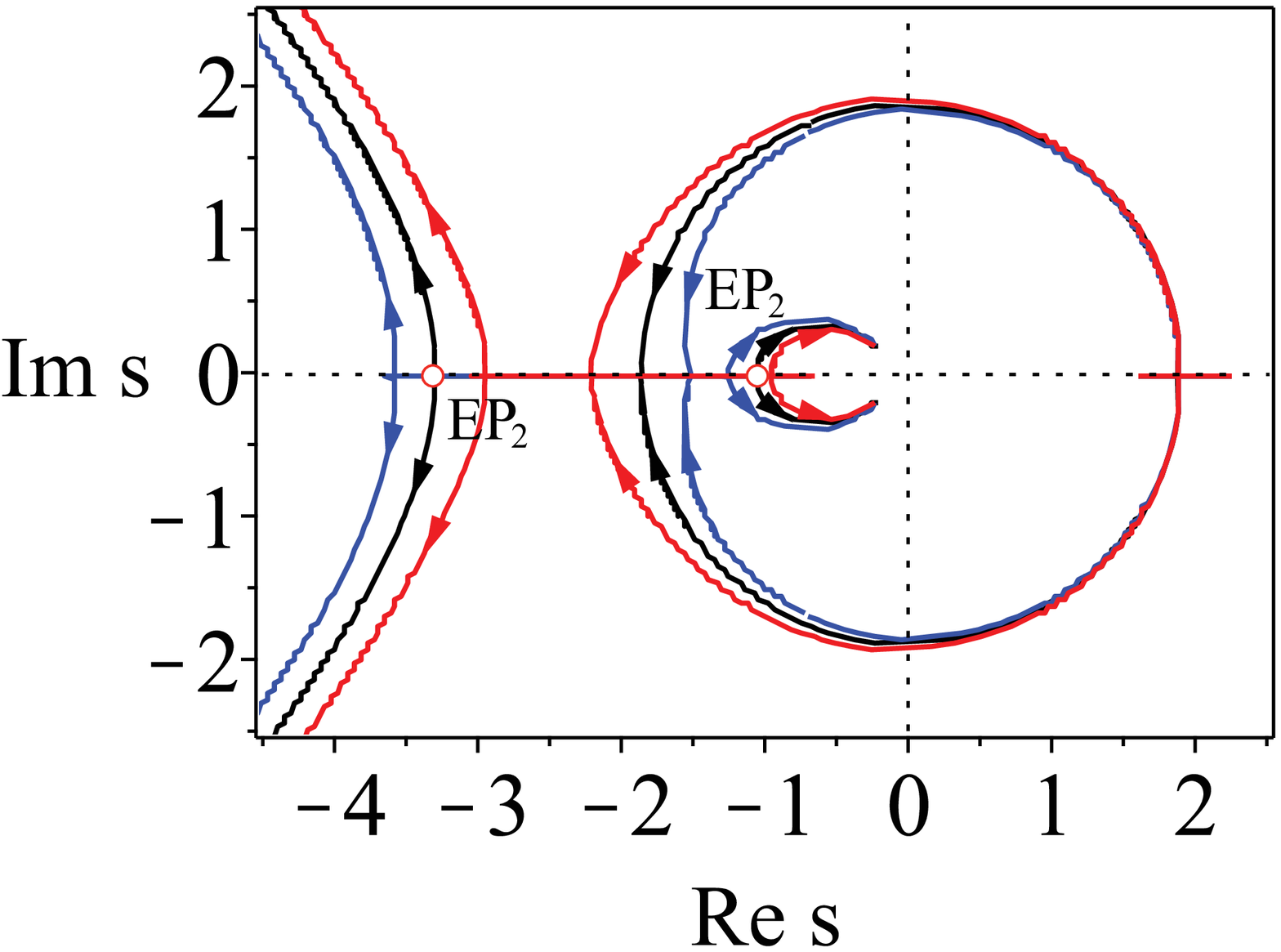}
    \includegraphics[angle=0, width=0.35\textwidth]{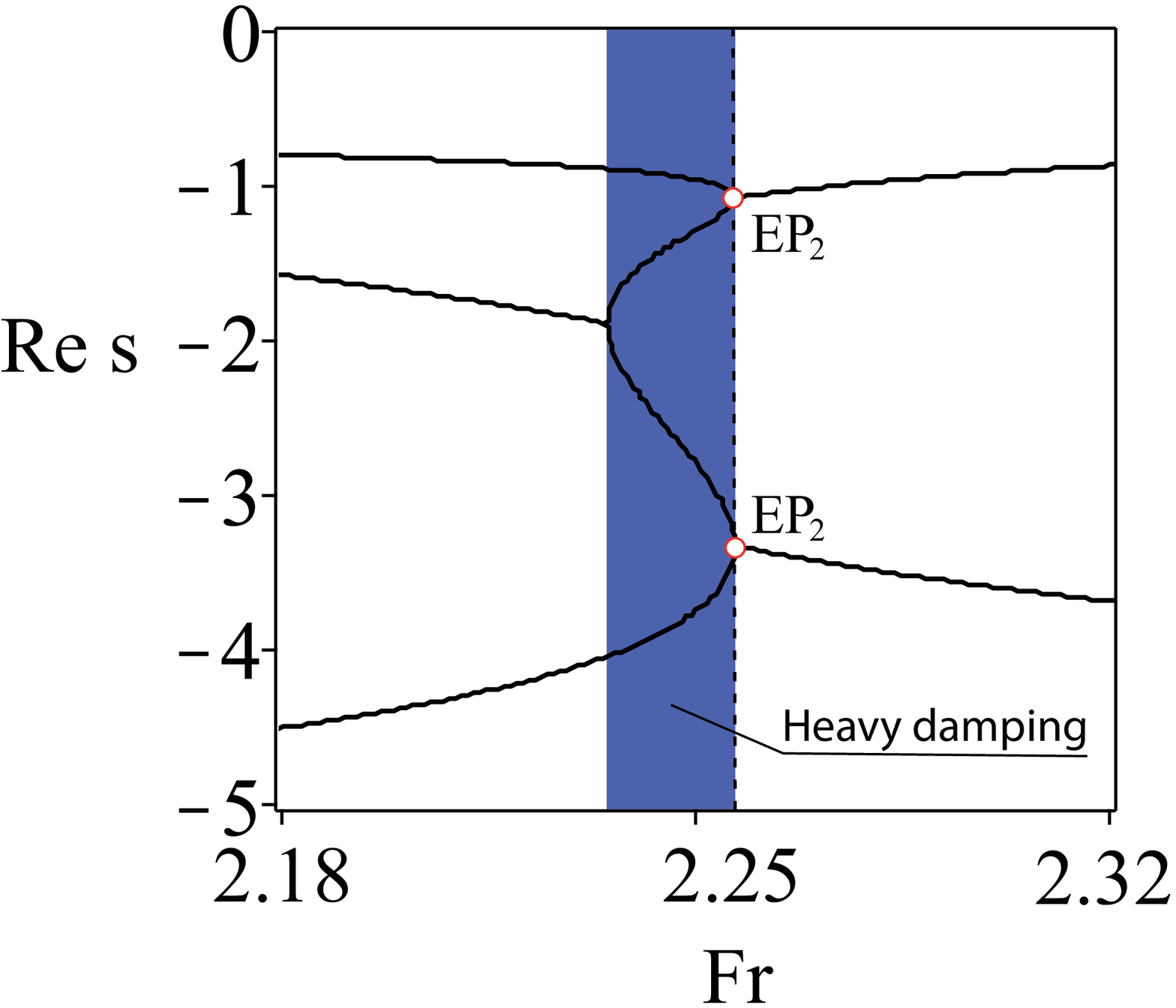}
    \includegraphics[angle=0, width=0.35\textwidth]{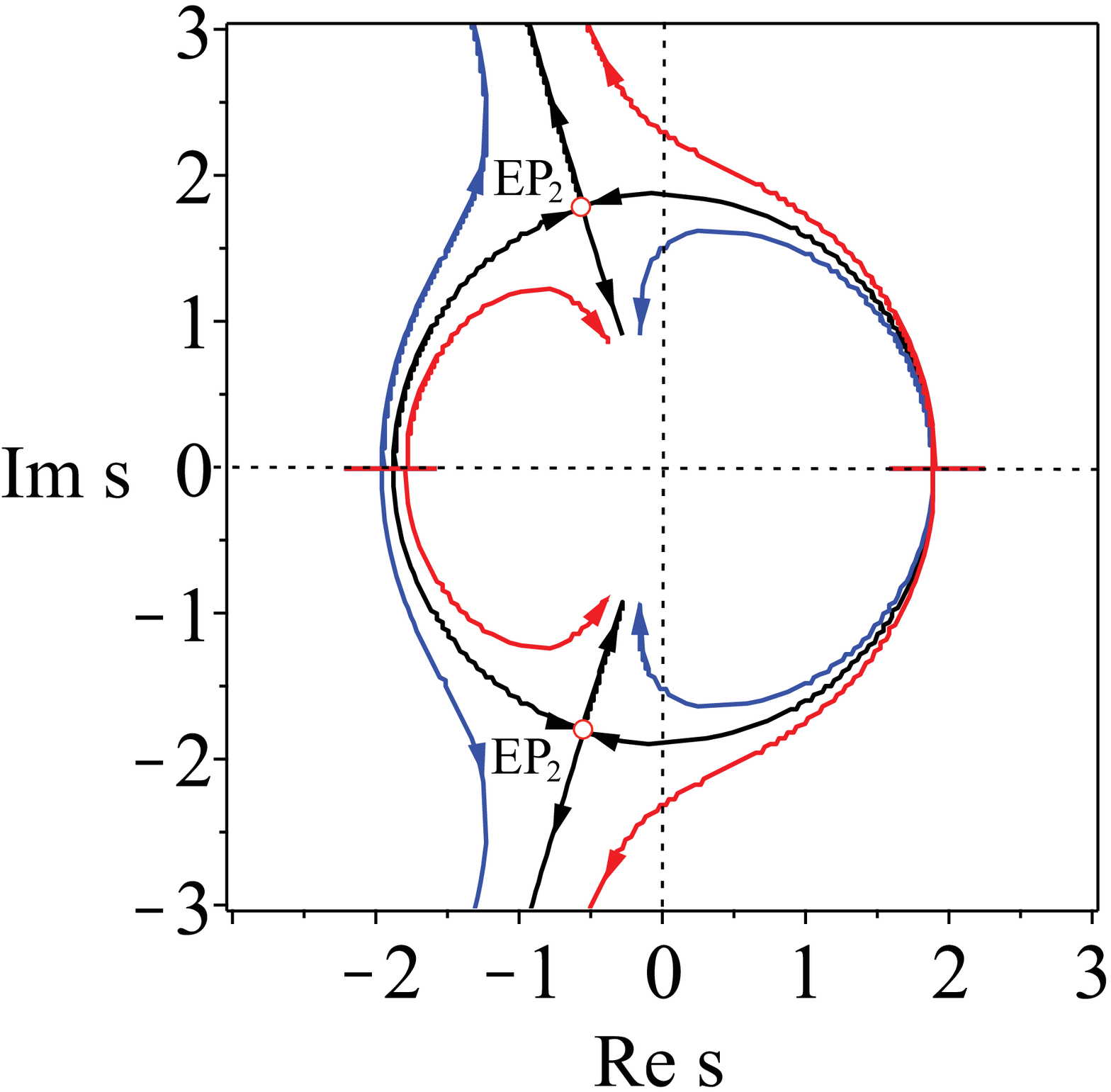}
    \includegraphics[angle=0, width=0.35\textwidth]{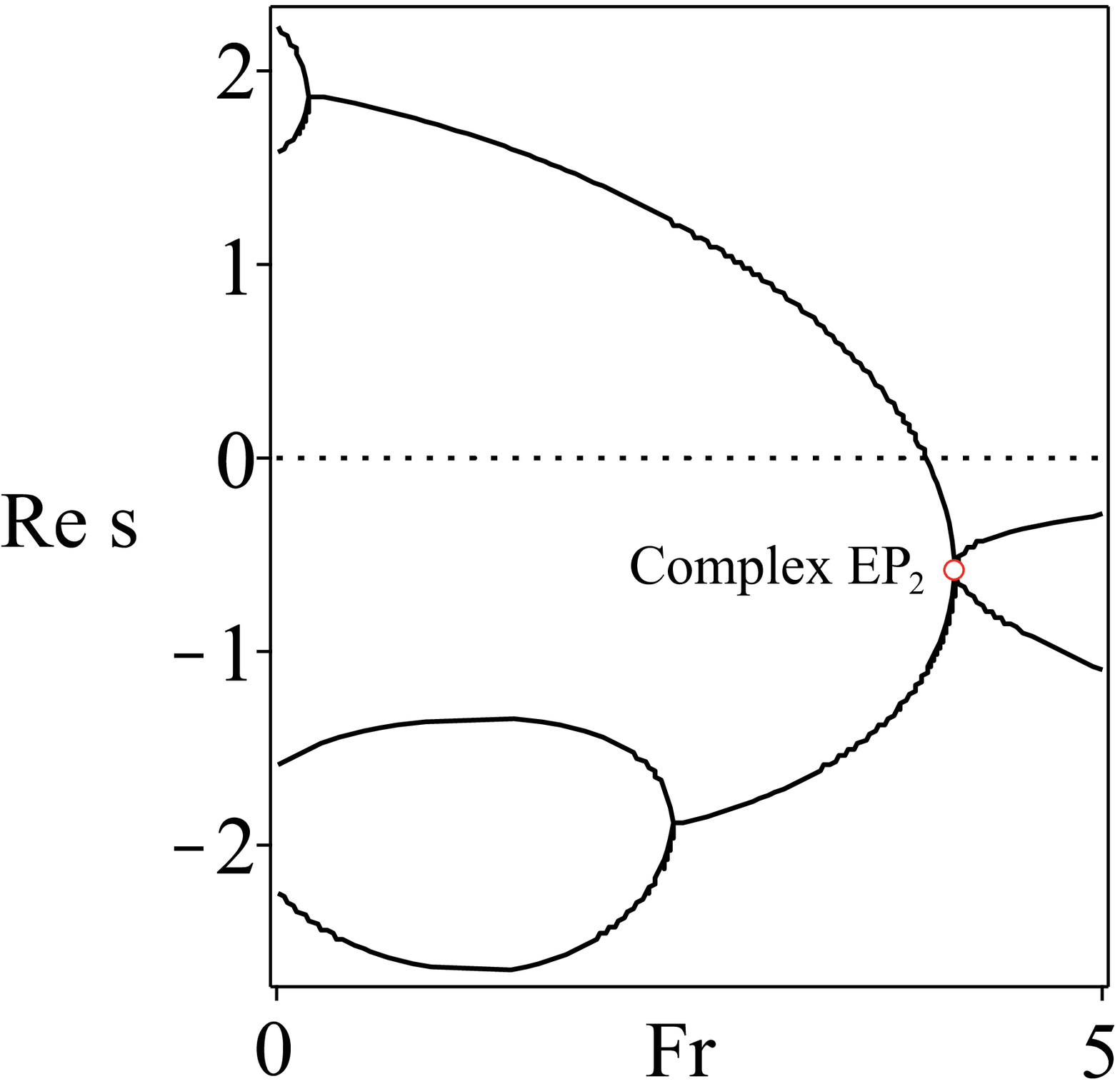}
    \end{center}
    \caption{$\chi_H=1$, $\zeta_H=-0.2$, $\zeta_B=-0.4$.
    Stabilization of the TMS bike as ${\rm Fr}$ is increasing from 0 to 5 for (upper row) $\lambda_s=\arctan\left(\frac{15}{4}-\frac{75}{32}\sqrt{2}\right)$ rad., (middle row)
    $\lambda_s=\arctan\left(\frac{15}{4}-\frac{75}{32}\sqrt{2}\right)-0.05$ rad., and (lower row) $\lambda_s=\arctan\left(\frac{15}{4}-\frac{75}{32}\sqrt{2}\right)+0.8$ rad. The eigenvalue curves are shown for
    (black) $\chi_B=\sqrt{2}$, (blue) $\chi_B=\sqrt{2}-0.01$, and (red) $\chi_B=\sqrt{2}+0.01$ in the upper and middle rows and for
    (black) $\chi_B=\sqrt{2}$, (blue) $\chi_B=\sqrt{2}-0.1$, and (red) $\chi_B=\sqrt{2}+0.1$ in the lower row. Notice the existence at $\chi_B=\sqrt{2}$ of (upper row) a real exceptional point ${\rm EP}_4$, (middle row) a couple of real exceptional points ${\rm EP}_2$, and (lower row) a couple of complex exceptional points ${\rm EP}_2$ and repelling of eigenvalue curves near ${\rm EP}$s when $\chi_B\ne\sqrt{2}$. }
    \label{fig3}
    \end{figure}

\subsubsection{Mechanism of self-stability and ${\rm CEP}_2$ as a precursor to bike's weaving}

What happens with the stability of bicycles that have large steer axis tilt? To answer this question let us look at the
movement of eigenvalues in the complex plane at different $\lambda_s$ and $\chi_B$ as the Froude number increases from 0 to 5, Fig.~\ref{fig3}.
At ${\rm Fr}=0$ the bicycle is effectively an inverted pendulum which is statically unstable ({\rm capsizing instability} \cite{Sharp2008}) with two real negative eigenvalues and two real positive eigenvalues. As $\rm Fr$ increases the positive eigenvalues move towards each other along the real axis. The same happens (at a slower rate) with the negative eigenvalues. Eventually, the positive real eigenvalues merge into a double real eigenvalue $s=-\omega_0>0$. The subsequent evolution of eigenvalues depends on $\chi_B$ and $\lambda_s$.

         \begin{figure}
    \begin{center}
    \includegraphics[angle=0, width=0.45\textwidth]{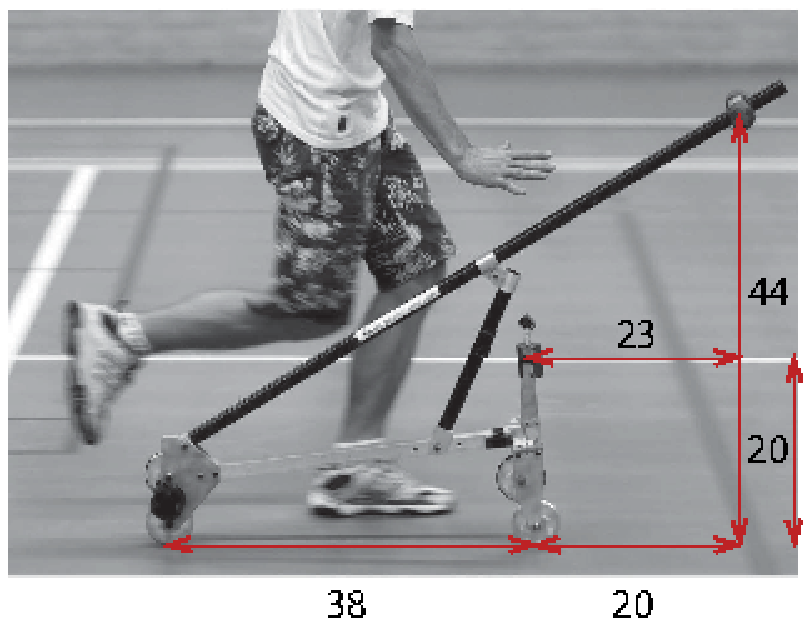}
    \end{center}
    \caption{Experimental realization of a self-stable TMS bicycle design found by trials and errors in \cite{TMS2011} with $\chi_B=1.526$, $\chi_H=0.921$, $\zeta_B=-1.158$, $\zeta_H=-0.526$ approximately fits the scaling law \rf{chib1}. Indeed, $\sqrt{\frac{\zeta_B}{\zeta_H}}=1.484$  is close to $\chi_B=1.526$. }
    \label{fige}
    \end{figure}

If $\chi_B=\sqrt{\frac{\zeta_B}{\zeta_H}}=\sqrt{2}$ then with the further increase in ${\rm Fr}$ the double eigenvalue $s=-\omega_0>0$ splits into a conjugate pair of complex eigenvalues with positive real parts causing weaving instability. This pair evolves along
a circle $({\rm Re} ~s)^2+({\rm Im} ~s)^2=\omega_0^2$ and crosses the imaginary axis exactly at ${\rm Fr}={\rm Fr}_c$ given by equation \rf{Frc1}, which yields the asymptotic stability of the bicycle.

The further evolution of the eigenvalues depends on the steer axis tilt $\lambda_s$, Fig.~\ref{fig3}. If $\lambda_s$ satisfies the constraint \rf{tls} then the complex eigenvalues with the negative real parts moving along the circle approach the real axis and meet the two negative real eigenvalues exactly at ${\rm Fr}={\rm Fr}_{{\rm EP}_4}$ forming a quadruple negative real eigenvalue $s=\omega_0$, i.e. the real exceptional point ${\rm EP}_4$. At this moment all the four eigenvalues are shifted as far as possible to the left from the imaginary axis, which corresponds to the minimum of the spectral abscissa, Fig.~\ref{fig3}(upper row). Further increase in ${\rm Fr}$ leads to the splitting of the multiple eigenvalue into a quadruplet of complex eigenvalues with negative real parts (decaying oscillatory motion).

If $\chi_B=\sqrt{\frac{\zeta_B}{\zeta_H}}=\sqrt{2}$ and $\lambda_s$ is smaller than the value specified by \rf{tls}, then the pair moving along the circle reaches the real axis faster than the negative real eigenvalues meet each other, Fig.~\ref{fig3}(middle row). Then, the complex eigenvalues merge into a double negative real eigenvalue $s=\omega_0$ which splits into two negative real ones that move along the real axis in the opposite directions. At these values of $\rm Fr$ the system has four simple negative real eigenvalues, which corresponds to heavy damping. The time evolution of all perturbation is the monotonic exponential decay, which looks favorable for the bike robustness. At ${\rm Fr}={\rm Fr}_{EP}$ which is determined by the equation \rf{epset} two real negative double eigenvalues originate simultaneously marking formation of the ${\rm 2EP}_2$ singularity on the boundary of the domain of heavy damping. Further increase in $\rm Fr$ yields splitting of the multiple eigenvalues into two pairs of complex eigenvalues with negative real parts (decaying oscillatory motion).

If $\chi_B=\sqrt{\frac{\zeta_B}{\zeta_H}}=\sqrt{2}$ and $\lambda_s$ is larger than the value specified by \rf{tls}, then the pair moving along the circle do it so slowly that the real negative eigenvalues manage to meet into a double negative real eigenvalue $s=\omega_0$ and then become a pair of two complex eigenvalues evolving along the same circle towards the imaginary axis, Fig.~\ref{fig3}(lower row). The pairs of complex eigenvalues meet on the circle at ${\rm Fr}={\rm Fr}_{\rm EP}$ which is determined by the equation \rf{epset}, i.e. at a point of the EP-set corresponding to a pair of complex exceptional points ${\rm EP}_2$. After the collision the eigenvalues split into four complex eigenvalues with the real parts.

From this analysis we see that $\lambda_s$ indeed determines the balance of the rate of stabilization of unstable modes and the rate of destabilization of unstable modes. The former is larger when $\lambda_s$ is smaller than the value specified by \rf{tls} and the latter is larger
when $\lambda_s$ exceeds the value specified by \rf{tls} thus confirming the design principle \rf{dp}. The perfect balance corresponds to the angle $\lambda_s$ specified by \rf{tls}, which yields minimization of the spectral abscissa.

When $\chi_B\ne \sqrt{\frac{\zeta_B}{\zeta_H}}$, then the eigenvalues evolve close to the circle $({\rm Re} ~s)^2+({\rm Im} ~s)^2=\omega_0^2$ but this evolution again differs for different values of $\lambda_s$. If for $\lambda_s$ smaller than the value specified by \rf{tls} the eigenvalue evolution remains qualitatively the same, as is evident from Fig.~\ref{fig3}(middle row), for  $\lambda_s$ larger than the value specified by \rf{tls} the eigenvalues experience strong repulsion near the location of ${\rm CEP}_2$, i.e. when the parameters evolve close to the EP-set of complex exceptional points. Such behavior of eigenvalues in dissipative systems attracted attention of many researchers. For instance, Jones \cite{Jones1988} wrote in the context of the stability of the plane Poiseuille flow that ``unfortunately, it is quite common for an eigenvalue
which is moving steadily towards a positive growth rate to suffer a sudden change of direction and subsequently fail to become unstable; similarly, it happens that modes which initially become more stable as [the Reynolds number] increases change direction and subsequently achieve instability. {\em It is believed that these changes of direction are due to the nearby presence of multiple-eigenvalue points.}'' This ``nearby presence'' of complex exceptional points is elusive unless we manage to localize the EP-set. For the TMS bike we have obtained this set in an explicit form given by equations \rf{w01}, \rf{chib1}, and \rf{epset}. Dobson et al. \cite{Dobson2001} posed a question ``is strong modal resonance a precursor to [oscillatory instability]?'' The strong modal resonance is exactly the interaction of eigenvalues at ${\rm CEP}_2$ shown in Fig.~\ref{fig3}(lower row). Knowing the exact localization of the EP-set of complex exceptional points we can answer affirmatively to the question of Dobson et al. Indeed, the complex EP-set shown as a dashed curve in Fig.~\ref{fig1}(left) tends to the boundary of asymptotic stability as $\lambda_s \rightarrow \frac{\pi}{2}$. This means that the ${\rm CEP}_2$ in Fig.~\ref{fig3}(lower row) come closer to the imaginary axis at large $\lambda_s$ and even small perturbations in $\chi_B$ can turn the motion of eigenvalues back to the right hand side of the complex plane and destabilize the system. Fig.~\ref{fig3}(lower row) also demonstrates the selective role of the scaling law $\chi_B=\sqrt{\frac{\zeta_B}{\zeta_H}}$ in determining which mode becomes unstable. The conditions $\chi_B>\sqrt{\frac{\zeta_B}{\zeta_H}}$ and $\chi_B<\sqrt{\frac{\zeta_B}{\zeta_H}}$ affect modes with the higher or the lower frequency, respectively. In fact, in the limit $\lambda_s \rightarrow \frac{\pi}{2}$ the dissipative system becomes close to a system with a Hamiltonian symmetry of the spectrum. This could be a reversible, Hamiltonian or PT-symmetric system \cite{K2012,K2013t,K2017} which is very sensitive to perturbations destroying the fundamental symmetry and therefore can easily be destabilized.

\subsubsection{How the scaling laws found match the experimental TMS bike realization}

In Fig.~\ref{fige} we show the photograph of the experimental TMS bike from the work \cite{TMS2011}. If we measure the lengths
of the bike right on the photo, we can deduce that for this realization the design parameters are $\chi_B=1.526$, $\chi_H=0.921$, $\zeta_B=-1.158$, $\zeta_H=-0.526$. Hence,
$$
\sqrt{\frac{\zeta_B}{\zeta_H}}=1.484   \approx \chi_B=1.526,
$$
which means that the scaling law \rf{chib1} is matched very well. This leads us to the conclusion that the trial-and-and error engineering approach to the design of a self-stable TMS bike reported in \cite{TMS2011} eventually has produced the design that is optimally stable with respect to at least three different criteria: minimization of the spectral abscissa, minimization of ${\rm Fr}_c$ and maximization of the domain of heavy damping. Indeed, our scaling laws \rf{chib1} and \rf{tls} directly follow from the exact optimal solutions to these problems.

\section{Conclusions}

We have found new scaling laws for the two-mass-skate (TMS) bicycle that lead to the design of self-stable machines. These scaling laws optimize stability of the bicycle by several different criteria simultaneously. The matching of the theoretical scaling laws to the parameters of the TMS bikes realization demonstrates that the trial-and-and error engineering of the bikes selects the most stable and thus empirically optimizes the bike stability. We have found the optimal solutions directly from the analysis of the sets of exceptional points of the TMS bike model with the help of a general result on the minimization of the spectral abscissa at an exceptional point of the highest possible order.
We stress that all previous results on stability of bicycles even in the linear case have been obtained numerically.

\vspace{6pt}


\section*{Acknowledgements}Support from a Vice Chancellor's
Research Fellowship at Northumbria University is gratefully acknowledged.





\bibliographystyle{mdpi}


\end{document}